\documentclass[11pt,centertags,leqno]{amsart}

\usepackage[foot]{amsaddr}
\usepackage{latexsym}
\usepackage[english]{babel}
\usepackage[T1]{fontenc}
\usepackage[numbers]{natbib}
\usepackage{amssymb}
\usepackage{fancyhdr}
\usepackage{url}
\usepackage{hyperref}
\usepackage{verbatim}
\usepackage{leftidx}
\usepackage{color,graphicx}
\usepackage{pdfpages}

\usepackage{mathrsfs, mathtools}
\usepackage{stmaryrd}
\usepackage{nicefrac}
\usepackage{marvosym}
\mathtoolsset{showonlyrefs}

\numberwithin{equation}{section} \makeatletter
\renewcommand{\subsection}{\@startsection
	{subsection}{2}{0mm}{\baselineskip}{-0.25cm}
	{\normalfont\normalsize\bf}} \makeatother

\theoremstyle{remark}

\def \A {\mathcal A}

\def \R {\mathbb R}

\def \bE {\mathbb E}

\def \X {\mathbf{X}}

\def \W {\mathbf{W}}

\newcommand{\ud}{\mathrm d}

\def \Nu {\boldsymbol{\nu}}

\begin{document}

	\author[R.~Frey]{R\"udiger Frey}\address{R\"udiger Frey, Institute for Statistics and Mathematics, Vienna University of Economics and Business, Welthandelsplatz, 1, 1020 Vienna, Austria }\email{rfrey@wu.ac.at}
	
	\author[T.~ Traxler]{Theresa Traxler}\address{Theresa Traxler, Institute for Statistics and Mathematics, Vienna University of Economics and Business, Welthandelsplatz, 1, 1020 Vienna, Austria }\email{theresa.traxler@wu.ac.at}

	\title[]{Playing with Fire? A Mean Field Game Analysis of Fire Sales and Systemic Risk under 
Regulatory Capital Constraints}

	\date{Preliminary version, \today}
	
	\begin{abstract}
We study the impact of regulatory capital constraints on fire sales and financial stability in a large banking system using a mean field game model. In our model banks adjust their holdings of a risky asset via trading strategies with finite trading rate in order to maximize expected profits. Moreover, a bank is liquidated if it violates a stylized regulatory capital constraint. We assume that the drift of the asset value is affected by the average change in the position of the  banks in the system. This creates strategic interaction between the trading behavior of banks and thus 
leads to a game. The equilibria  of this game are characterized by a system of coupled
PDEs. We solve this system explicitly for a test case without regulatory constraints and
numerically for the regulated case. We find that capital constraints can lead to a systemic crisis where a substantial proportion of the banking system defaults simultaneously.  Moreover, we discuss  proposals from the literature on macroprudential regulation. In particular, we show  that in our setup a systemic crisis does not arise if the banking system is sufficiently well capitalized or if improved mechanisms for the resolution of banks violating the risk capital constraints are in place.  


	\end{abstract}
	
	\maketitle
	
	{\bf Keywords}: Mean field game , Systemic risk, Price-mediated contagion, Risk capital constraints,  McKean-Vlasov equation


\section{Introduction}

Contagious interactions between financial institutions play an important role in the amplification of economic shocks during a financial crisis.  A prime example is the global  financial crisis of 2007-2009, where comparatively small losses on the market for US subprime mortgages were magnified by the financial system and caused a major recession whose repercussions were felt throughout the globe. This has led to a large literature on financial contagion and systemic risk. Most of this work focuses on \emph{direct contagion}  generated by contractual links between financial institutions such as interbank lending or OTC derivatives. Examples of this line of research include \citet{bib:eisenberg-noe-01}, \citet{bib:elsinger-lehar-summer-06}, \citet{bib:rogers-veraart-13}, \citet{bib:glasserman-young-15} or \citet{bib:frey-hledik-18}.
Indirect or \emph{price-mediated contagion} on the other hand  is caused by price effects due to forced  de-leveraging, where a distressed financial institution is rapidly selling some of its risky assets in order to stay solvent or to comply with regulatory constraints  in so-called \emph{fire sales}.  These fire sales tend to exhaust demand for  risky assets, which pushes asset   prices downward and possibly   leads to losses of other financial institutions as well. The relevance of price-mediated contagion is stressed in many reports and policy papers  on the great financial crisis, see    \citet{bib:hanson-kashyap-stein-11} or the introduction of \citet{bib:braouezec-wagalath-19}. In  particular, in \cite{bib:basel-14} the Basel committee acknowledges  that
\begin{quote} at the height of the crisis financial markets
forced the banking sector to reduce its leverage in a manner that amplified downward pressures on asset
prices. This de-leveraging process exacerbated the feedback loop between losses, falling bank capital and
shrinking credit availability.
\end{quote}
According to the literature on macroprudential regulation such as \citet{bib:hanson-kashyap-stein-11}, \emph{regulatory capital constraints} may be an important contributor  to price-mediated contagion:  under the Basel capital adequacy framework  the risk capital of a bank (its equity and certain forms of long-term debt) must exceed a multiple of its  \emph{risk weighted assets}   (a measure of the overall size of the risky assets of a bank scaled by their relative riskiness). If a bank incurs substantial losses, its risk capital might be reduced tho the extent that it no longer complies with the Basel rules. Since in such a situation it is usually too costly to issue new equity, the bank is forced to engage in  fire sales  in order to  avoid regulatory penalties or  liquidation. In a recession scenario where many banks experience losses simultaneously, these fire sales might destabilize the banking system via price mediated contagion.   

Despite the relevance of the topic there are only  few mathematical models that analyze the implications  of regulatory capital constraints for  systemic risk, see the literature review below. 
In this paper we analyze price-mediated contagion in the context of a  mean field game  model
for a large banking system. In our setup a bank invests  into a non-tradable asset, for instance retail loans, into a  tradable but illiquid risky asset, for instance tradable credit positions  and in cash,  with the  objective of   maximizing   the  expected value of its equity  at some horizon date $T$. The position in the tradable  asset can be adjusted only gradually, so that we confine  banks to  trading strategies that are absolutely continuous with a finite  \emph{trading rate}.   Moreover,  banks incur transaction costs which depend on the size of the trading rate,  making  a very rapid adjustment of their position prohibitively costly.   We assume that  the drift of the tradable assets depends on the change in the average asset holdings  of the  banks in the system. In particular prices are pushed downward  if the banking sector reduces its overall position in the tradable assets, so that there is  price-mediated contagion.  The optimal trading rate of a bank obviously  depends on the drift of the risky asset and hence on the trading strategies of the other banks, so that there is interaction between their trading strategies.  In mathematical terms we are dealing with a so-called mean-field game of controls (MFGC), where agents interact via the distribution of their states and  via the distribution of the controls (in our case their trading strategies).  
We introduce a stylized risk capital constraint and impose the condition that the equity value of a bank needs to exceed a multiple of its asset holdings at all times. Upon violation of this condition  a  bank is liquidated by regulators and  its assets are sold on the market. 

In the case without capital constraints the mathematical structure of our model  is similar to the MFGC used  by  \citet{bib:cardaliaguet-leHalle-18} to study strategic interaction in  optimal portfolio execution. In particular, the PDE system for the game (the dynamic programming equation for the control problem of individual banks and the forward equation for the distribution of bank characteristics) can be reduced to a system of ODEs that has an explicit solution. In the case with capital constraints, on the other hand,  the PDE system for the mean field game becomes a boundary value problem for which there is no explicit solution. Hence  we have to resort to numerical methods.  The numerical analysis of the PDE system corresponding to a MFG is challenging since one deals simultaneously  with a forward and a backward equation and hence with a fixed point problem. For our MFGC model we use Picard iterations combined with a special discretization scheme that uses some ideas from \citet{bib:achdou-20}.  

Using this scheme we study numerically the impact of capital constraints case on the stability of the banking system. We find that  banks liquidate their position faster than in the unregulated case,  in particular when they are close to the liquidation boundary, and that they hold on average a lower amount of the risky asset, so that  the risk bearing capacity of the banking system is reduced by the capital constraints.  For certain parameter values we observe a \emph{liquidation cascade}  where many banks are violating the risk capital constraints more or less simultaneously and where the ensuing fire sales have a substantial impact on the drift of the risky assets.
These findings  lends formal support to the conjecture from the regulatory literature  that risk capital constraints may cause price-mediated contagion. With this in mind we test several macroprudential risk management  policies   (regulatory measures  trying to mitigate the external effects of rapid deleveraging by multiple financial institutions) from the regulatory literature, see \citet{bib:hanson-kashyap-stein-11}).
In particular, we find  that in our setup  the adverse  effects of capital constraints disappear  if  the banking system is sufficiently well capitalized. Moreover, we show that financial stability is enhanced if banks violating  the  capital constraints are resolved in a way that mitigates  price-mediated contagion, for instance by parking their assets  in a special purpose vehicle that is unwound only gradually over time.  

The remainder of the paper is organized as follows. In Section~\ref{sec:model} we introduce our setup and the optimization problem of an individual bank. The case without capital constraints, where the MFGC has an explicit solution, is studied in Section~\ref{sec:unregulated}. In Section~\ref{sec:regulated} we discuss the PDE system for the MFGC model with capital constraints. Numerical experiments studying the implications of capital constraints for financial stability are discussed in Section~\ref{sec:numerical_analysis}.

\subsection*{Literature review.}
We begin with the literature that studies the  implications of regulatory capital constraints in formal economic
models. \citet{bib:braouezec-wagalath-19} consider a one period model with finitely many banks that hold positions in the same risky asset. They  assume that the value of this asset is hit by an  exogenous shock at time $t=1$ and that every  bank sells the minimal amount of assets necessary to comply with the Basel rules. The market for the asset is not perfectly liquid. Hence the  fire sales of one bank create an externality for the other banks in the system and the liquidation problem becomes a game.   \citet{bib:braouezec-wagalath-19} prove  that this game  has at least one  equilibrium. Moreover, they run simulation experiments  with input data calibrated to the American banking system, which show that in their model price mediated contagion  significantly amplifies the impact of the initial  shock.  \citet{bib:feinstein-21} extends this analysis to a setting in continuous time with deterministic asset prices. We stress that in these models banks react in a quite mechanical way to the risk capital constraints. In our model on the other hand, risk capital constraints and the ensuing price-mediated contagion  are factored into the dynamic investment strategies of the banks and influence their behavior prior to reaching the liquidation boundary. 
For further work on fire sales and price-mediated contagion we refer to  \citet{bib:cont-wagalath-16} or  \citet{bib:cont-schaaning-19} and the references therein.  

Next we discuss  mean field game models for systemic risk. A simple model is proposed  in  \citet{bib:carmona-fouque-15}. More relevant for our work are  the  papers by \citet{bib:nadtochiy-shkolnikov-19}, \citet{bib:hambly-ledger-sojmark-19}, \cite{bib:hambly-sojmark-19} or \citet{bib:cuchiero-rigger-svaluto-22}. These papers extend the neuron-firing model of \citet{bib:delarue-et-al-15} to systemic risk. More precisely, they study the mean field limit of banking  systems where every bank suffers a loss if one bank in the system crosses an exogeneous default barrier. 
Fundamental contributions on mean field games include   \citet{Lasry2006a,Lasry2006b} and \citet{carmona2018}. \citet{Lasry2006a,Lasry2006b}  focus on  the PDE approach (as we do in the present paper) whereas  \citet{carmona2018}  discuss  the probabilistic approach based on forward backward SDEs. A good general introduction to the topic is given in  \citet{bib:cardaliaguet-porretta-19}.

\section{The Model}\label{sec:model}

\subsection{The Banking System}
Fix some horizon date $T$. We consider a large banking system with a continuum $I$ of  stylized banks.  Each bank  $i\in I$ holds some non-tradable asset with value  $A^i$, for instance retail loans.  Moreover,  it invests into a risky asset $S^i$, for instance tradable credit positions,  and in cash.  We assume that the market for the asset $S^i$ is not perfectly liquid so that a bank can adjust its position only gradually over time. In mathematical terms it is restricted to  trading strategies  with finite trading rate $\Nu^i= (\nu_t^i)_{0\le t \le T}$.  Moreover,  there are transaction costs that are proportional to $(\nu_t^i)^2$; in this way we penalize a very rapid change of positions.    We denote  by   $Q^i = (Q_t^i)_{0 \le t \le T} $ the amount of the tradable risky assets held by bank $i$  (also referred to as \emph{inventory level}) and by $C^i = (C_t^i)_{0 \le t \le T}$ its cash position. We allow for $C_t^i <0$  (this corresponds to the case where the bank is a net borrower) and we work for simplicity with an interest rate $r=0$.

For a given trading strategy $\Nu$, the the dynamics of $(A^i,Q^i,S^i) $ are
\begin{align}\label{dynamics}
\ud A_t^i &= \sigma_A \ud W_t^{A,i}\\
\ud Q_t^i &= \nu_t^i \ud t + {\sigma_Q \ud {W}_t^{Q,i}}\\
\ud S_t^i  &= (\mu_{\text{ex}} + \alpha \overline{\mu}_t) \ud t + \sigma_S \ud W_t^i
\end{align}
for parameters $\alpha,  \sigma_A, \sigma_S, \sigma_Q  \ge 0$ and a three-dimensional standard  Brownian motion  $\W^i= (W^{Q,i}, W^{A,i}, W^{S,i})$. The transaction cost incurred by bank $i$ is given by   $\kappa (\nu_t^i)^2$ for some $\kappa >0$. We assume that the trading strategies used by bank $i$ will be  adapted to the filtration generated by the Brownian motion $\W^i$ and  sufficiently integrable, that is $\int_0^T \left(\nu_s^i\right)^2 \ud s < \infty$. Moreover, the Brownian motions are independent across banks. 
The cash account is used to finance the trading activities of the bank and it collects proceeds from asset sales;  its dynamics are thus given by
$$\ud C_t^i  = - S_t^i \ud Q_t^i - \kappa \left(\nu_t^i\right)^2 \ud t.$$
Denote by
\begin{equation}
X_t^{i,\Nu}  = A_t^i +S_t^i Q_t^{i,\Nu} + C_t^{i,\Nu}  \label{eq:equity}
\end{equation}	
the  value of the equity of bank $i$ given that it uses the  trading strategy $\Nu$.  Banks choose their trading rate $\Nu$ in order to maximize their  expected equity value at the horizon date $T$, $\bE \big ( X_T^{i,\Nu} \big)$,  over  all  strategies $\Nu$.
For comparison purposes, in the case without capital constraints we also  consider the more general problem of maximizing $\bE \left ( X_T^{i,\Nu} - \gamma \big(Q_T^{i,\Nu}\big)^2 \right )$ for some $\gamma >0$, where the  penalty $\gamma (Q_T^{i,\Nu})^2  $ can be interpreted as a terminal liquidation cost.

\subsubsection*{Comments.} (i) The drift of $S^i$ depends on two components: $\mu_{\text{ex}}$ represents some exogenous trend and the \emph{interaction} or \emph{contagion term} $\overline{\mu}_t$ is the rate of change in the average number of risky assets held by the banking sector  (a precise definition is given in \eqref{eq:def-mu_bar} and \eqref{eq:interaction-term-regulated} below). In particular the drift of $S^i$ decreases if the banking sector reduces its overall position in the tradable assets. Note that  the optimal trading rate of bank $i$ depends on the drift of $S^i$ and hence on  the contagion term $\overline{\mu}_t$, which in turn depends  on the  trading strategies of the other banks in the system. Hence there is strategic interaction between the trading of the banks and  are dealing with a game.

(ii) The diffusion term in the dynamics of $Q^i$ reflects the fact that large banks are usually not able to perfectly control the exact amount of risky assets they hold, for instance due to  execution delays. From a mathematical  perspective, the diffusion term in the dynamics of $A^i$ and $Q^i$ ensures that  the PDE system characterizing the mean-field game is  uniformly parabolic.

(ii) In our model the Brownian motions $\W^i$, $i \in I$, are independent, so that  banks interact only via the impact of their trading on the drift $\overline{\mu}$. It might make sense to consider a common noise component in the asset price, but this more involved case  is left for future research.

\subsection{The HJB equation}
Next we derive the HJB equation for the optimization problem of a generic  bank. From now on we omit the superscript $i$ since the banks in our system are all identical.
Using Itô's product formula and the assumption that Brownian motions are independent  we get  the following dynamics for the process $\mathbf{X} =(Q,X)$,
\begin{align*}
    \ud Q_t &= \nu_t \ud t + \sigma_Q \ud W_t^Q, \\
    \ud X_t &= \ud A_t + \ud (S_tQ_t ) + \ud C_t = \ud A_t + Q_t\ud S_t + S_t\ud Q_t -  S_t \ud Q_t - \kappa \nu_t^2 \ud t  \\
        &= \left( Q_t( \mu_{\text{ex}} + \alpha \overline{\mu}_t )- \kappa \nu_t^2 \right) \ud t + \sigma_A \ud W_t^A  + Q_t \sigma_S \ud W_t^S.
\end{align*}
Assume furthermore  that $\overline{\mu}_t$  is given by a known  function $\mu(t)$ ($\mu(t)$  can  be interpreted  as a bank's expectation of the  contagion  term.)   It follows that for a constant strategy $\nu_t \equiv \nu$ the process $\mathbf{X}$ is Markov with generator
\[
 \mathcal{L}^{\mathbf{X},\nu} f = \nu \partial_q f + \left(q( \mu_{\text{ex}} + \alpha \mu(t) )- \kappa \nu^2\right) \partial_x f + \frac{1}{2} \sigma_Q^2 \partial_q^2 f + \frac{1}{2} \left(\sigma_A^2 + \sigma_S^2 q^2 \right) \partial_x^2 f . \]
Standard arguments of stochastic control theory give the following HJB equation for the value function $u$  of the bank's optimization problem
\[0= \partial_t u + \sup_{\nu \in \R} \left\{ \nu \partial_q u + \left ( q( \mu_{\text{ex}} + \alpha \mu)- \kappa \nu^2\right ) \partial_x u + \frac{1}{2} \sigma_Q^2 \partial_q^2 u + \frac{1}{2} \left(\sigma_A^2 + \sigma_S^2 q^2 \right )\partial_x^2 u  \right\}, \]
or, equivalently,
\begin{equation}
\begin{aligned}
    0&= \partial_t u + q( \mu_{\text{ex}} + \alpha \mu(t) ) \partial_x u + \frac{1}{2} \sigma_Q^2 \partial_q^2 u + \frac{1}{2} \left(\sigma_A^2 + \sigma_S^2 q^2 \right) \partial_x^2 u
    + \sup_{\nu \in \R } \left\{ \nu \partial_q u  - \kappa \nu^2 \partial_x u  \right\} \,;
\end{aligned}
\label{HJB_bankingsystems}
\end{equation}
the terminal condition is $u(T,q,x) = x -\gamma q^2.$ Note that in an equilibrium of the game  the predicted drift $\mu(t)$ and the realized drift $\overline{\mu}_t$ should coincide so that we get the additional  consistency or equilibrium condition $\mu(t) = \overline{\mu}_t$.

\subsection{Risk capital constraints}
Under the Basel capital adequacy rules  the risk capital of a bank  must exceed  a multiple of its so-called \emph{risk weighted assets} (RWA). Loosely speaking, this quantity is a weighted average of the size of the position in different asset classes, where the weights reflect differences in riskiness.  In this paper we introduce a stylized version of this constraint. We model the risk capital of a bank at time $t$ by the book value of its equity $X_t$ and the risk weighted assets by the term $\tilde \beta |Q_t| + \tilde c$ where  $\tilde \beta |Q_t|$ represents the risk of the position in the tradable asset $S$ and the  constant $\tilde c$ the risk of the position in the non-traded asset. This leads to a condition of the form
$
 X_t  > \beta |Q_t| + c
$ for constants  $c, \beta >0$.  We denote by the open set $\mathcal{A}$ the  set of \emph{acceptable positions}, that is
\begin{equation} \label{acceptable}
\mathcal{A} =\{\mathbf{x} = (q,x) \in \R \times \R^+ \colon x > \beta |q| + c \}
\end{equation}
and we we denote its boundary by $\partial \A$. In the case with capital constraints we assume that a bank  is  liquidated by the regulator at the stopping time  $\tau = \inf \{t \ge 0 \colon \X_t \notin \mathcal{A}\}$, that is as soon as  its position  reaches $\partial \A$.   We  assume that the equity holders lose their claim to the bank's equity in that case. This leads to the  boundary condition $u \equiv 0$ on $\partial \A$. (For mathematical reasons, in   Section~\ref{sec:regulated} we will  consider a slightly relaxed version of this condition.) In the sequel we refer to a bank with $\tau >t$ as an \emph{active} bank at $t$.

\section{The Case without Capital Constraints} \label{sec:unregulated}

In the case without capital constraints (the so-called unregulated case)  we denote the value function by $u^{\text{unreg}}$. Choosing a similar approach as \citet{bib:cardaliaguet-leHalle-18}, we obtain an explicit solution for $u^{\text{unreg}}$ and for the optimal strategy. 
We make the Ansatz $u^\text{unreg}(t,q,x)=x+v(t,q)$. This implies that  $\partial_x u^\text{unreg} =1$ and we get the following HJB equation for $v$ (for a given interaction term  $\mu(t)$)
\begin{equation} \label{eq:HJB-for-v}
 0= \partial_t v + q(\alpha \mu(t)  + \mu_{\text{ex}})
+ \sup_{\nu \in \R} \left\{ \nu \partial_q v - \kappa \nu^2  \right\} ,
\end{equation}
with terminal condition $v(T,q)=- \gamma q^2$. It follows that the optimal strategy is given by
\begin{equation} \label{eq:strategy-unreg-1}
\nu^* (t,q) = \frac{\partial_q v(t,q)}{2 \kappa},
\end{equation}
in particular, $\nu^*$ is independent of $x$. To find an explicit solution for $v$ we make the Ansatz
\[v(t,q)=h_0(t) + h_1(t)q- \frac{1}{2}h_2(t)q^2.\]
Note that $\sup_{\nu \in \R} \left\{ \nu \partial_q v  - \kappa \nu^2  \right\} =  \frac{(\partial_q v)^2}{4 \kappa}$. Substituting the Ansatz for $v$ into the HJB \eqref{eq:HJB-for-v} gives
\[ 0= h_0' + h_1' q - \frac{1}{2} h_2' q^2 + q(\alpha \mu(t) + \mu_{\text{ex}})
+ \frac{h_1^2-2h_1h_2q+ h_2^2q^2}{4 \kappa}, \]
which yields the following ODE system for $h_0,h_1,h_2$
\begin{align*}
    h_2' &= \frac{h_2^2}{2 \kappa} ,\quad
    h_1' = - \alpha \mu(t) - \mu_{\text{ex}} + \frac{h_1 h_2}{2 \kappa} ,\quad
    h_0' = - \frac{h_1^2}{4 \kappa} 
\end{align*}
with terminal conditions $h_0(T)=h_1(T)=0$ and $h_2(T)=2\gamma$.
There is an explicit solution of the ODE for $h_2$,
\begin{equation} \label{eq:h_2}
\begin{cases}
    h_2(t)= \frac{2 \kappa}{ T-t+ \frac{\kappa}{\gamma}} & \gamma>0, \\
    h_2(t) \equiv 0 & \gamma=0.
\end{cases}
\end{equation}
Moreover, the optimal strategy is given by $\nu^*(t,q) = \frac{1}{2 \kappa} (h_1(t)-h_2(t)q)$. Note that this strategy is linear in $q$ (and in fact even constant if $\gamma =0$).

Since $\nu^*$ depends only on $q$ (and not on the equity value $x$), each agent is fully characterized by his current inventory, so that we may describe the distribution of agents in terms of the distribution of inventory levels. We denote  the distribution of $Q_t$ $m_t(\ud q)$ and we  use for $f:\mathbb{R} \to \mathbb{R}$ the notation $\langle m_t,f \rangle: =\int_\mathbb{R} f(q) m(t,\ud q)$.
Recall that the interaction term in the drift is of the assets is  the change in the average inventory level. Using the above notation we get  the following formal definition of  $\overline \mu_t$
\begin{equation}\label{eq:def-mu_bar}
    \overline \mu_t = \int_\mathbb{R} q\,  m(t,\ud q) = \langle m(t), q\rangle\,.
\end{equation}

Denote by $\mathcal{L}^Q$ the generator  of $Q$. For $f$ in the domain of $\mathcal{L}^Q$ the weak form of the {forward}  equation for the evolution of $m_t (\ud q)$ reads
\[ \partial_t \langle m_t,f\rangle = \langle m_t,\mathcal{L}_t^Q f \rangle\,. \]
If $m_t(\ud q)$ has a density, i.e. $m(t,\ud q)=m(t,q)\ud q$ for all $t$, partial integration gives the  classical forward equation for $m(t,q)$  (but in the case without capital constraints we do not need this equation).
Using the forward equation and the fact that $ \mathcal{L}^Q q  =\nu^*$ we get that
\begin{equation}\label{eq:ODE-E}
 \overline \mu_t =  \partial_t \langle m_t,q \rangle = \langle m(t),\mathcal{L}_t^Q q \rangle = \langle m_t,\nu^*(t,\cdot) \rangle \,,
\end{equation}
so that $\overline{\mu}$ can  be interpreted as average trading rate of the banks. 

Following \cite{bib:cardaliaguet-leHalle-18}, we introduce the average inventory level $E(t)= \langle m(t), q\rangle $. It follows from \eqref{eq:def-mu_bar} that the contagion  term  $\overline \mu_t$ equals $E'(t)$. 
Using  \eqref{eq:ODE-E} and the  relation $\nu^*(t,q) = \frac{1}{2 \kappa} (h_1(t)-h_2(t)q)$ we thus get  the following ODE for the average inventory $E(t)$
$$
E'(t)= \langle m_t,\nu^*(t,\cdot) \rangle = \frac{1}{2\kappa} \left( h_1(t) - h_2(t) E(t) \right)\,,
$$
with initial condition $E_0 = \langle m_0, q\rangle$, where $m_0( \ud q)$ is the given initial distribution of the inventory $Q$. Using finally the equilibrium condition $\mu(t) = \overline{\mu}_t$, we get the following ODE system for $h_0,h_1$ and $E$ ($h_2$ is given in \eqref{eq:h_2})
\begin{subequations}
\label{eq:system}
\begin{align}
    h_1' &= -\frac{\alpha }{2\kappa}\left( h_1 - h_2 E \right) - \mu_{\text{ex}} + \frac{h_1h_2}{2\kappa}  & h_1(T)=0\,, \label{eq:b} \\
    h_0' &= -\frac{h_1^2}{4 \kappa}  & h_0(T)=0\,, \label{eq:c} \\
    E' &= \frac{1}{2\kappa} \left( h_1 - h_2 E \right)  & E(0)=E_0. \label{eq:d}
\end{align}
\end{subequations}
The ODE system \eqref{eq:def-mu_bar} can be viewed as a reduced form of the PDE system that usually describes the equilibrium in a MFG. In fact, the backward equation~\eqref{eq:b} derives from the HJB equation and it ensures that trading rate of a bank is optimal given  the interaction term whereas \eqref{eq:d} comes from the forward equation for the distribution of the inventory. 

Next we extend the arguments from \cite{bib:cardaliaguet-leHalle-18} to obtain an explicit solution.  We get from \eqref{eq:d} that $h_1 = 2\kappa  E' + h_2E, \label{h1_eq}$ and hence $h_1' = 2\kappa E'' + h_2' E + h_2 E'.$
Plugging the second relation into \eqref{eq:b} yields
\begin{align*}
    0  &= - (2\kappa E'' + h_2' E + h_2 E') - \alpha E'  - \mu_\text{ex} +  E'h_2 + \frac{  h_2^2E}{2\kappa} \\
    &= - 2\kappa E'' - \alpha E'  + \frac{  h_2^2 - 2\kappa h_2'}{2\kappa} E - \mu_\text{ex} .
\end{align*}
By \eqref{eq:h_2}, we get the following second order ODE for $E$
\begin{equation}
    E''+ \frac{\alpha}{2\kappa} E' = -\frac{\mu_\text{ex}}{2\kappa},
    \label{E_ODE}
\end{equation}  
with initial condition $E(0)=E_0$ and with $E(T)$ solving $E'(T) + \frac{\gamma}{\kappa}E(T)=0$ (this  follows from \eqref{eq:d}).
Solving this problem yields 
\[ E(t) = \frac{ E_0 \left( \alpha^2 e^{-\frac{\alpha}{2\kappa}T} + 2 \gamma \alpha \left(  e^{-\frac{\alpha}{2\kappa}t} - e^{-\frac{\alpha}{2\kappa}T}\right) \right) - \left( 2 \kappa \mu_{\text{ex}} + 2 \mu_{\text{ex}} \gamma T\right) \left( e^{-\frac{\alpha}{2\kappa}t} -1 \right)   }{ \left( \alpha^2 -2\gamma \alpha \right) e^{-\frac{\alpha}{2\kappa}T} + 2\gamma \alpha }  - \frac{\mu_{\text{ex}}}{\alpha} t  .\]
Given $h_2$ and $E$, we can compute $h_1$, $h_0$ and therefore also $v$ and $\nu^*$ by integration. The corresponding formulas are given in Appendix~\ref{app:explicitSol}.

\section{The case with capital constraints} \label{sec:regulated}

\subsection{The PDE system for the MFGC} 

In the case with capital constraints  the Ansatz  $u(t,q,x) = x + v(t,q)$  is inconsistent with the boundary condition for $u$ on  $\partial \A$. Moreover,  we expect that the  optimal trading rate  depends both on $x$ and $q$, since  close to the boundary of  $\mathcal{A}$ the bank will want to reduce its position in order to avoid liquidation. Hence we  need to work with the two-dimensional state price process $\mathbf{X} =(X,Q)$ and with the full  HJB equation \eqref{HJB_bankingsystems}.  

To complete the description of the HJB equation in the case with capital constraints we next give a precise description of the terminal and boundary condition we impose on $u$. At $T$ we assume that $u(T,x,q) = x$. Consider next for fixed $\epsilon >0$   the function $k\colon [0,T] \to \R$ with
  \[k(t)=\frac{1}{2 \epsilon} \left( t-T+\epsilon + \sqrt{0.0004+(t-T+\epsilon)^2} \right), \]
and note that $k$ is a smooth function with the following properties: $k(T) \approx 1$; $k(t)$ is non-decreasing for all $t \leq T$; $k(t)\approx 0$ for all $t \leq T-\epsilon$.
In the following we assume that
\begin{equation} \label{eq:boundary-cond-u}
 u (t,q,x) = k(t) \cdot (\beta \lvert q \rvert +c) , \quad (t,q,x) \in [0,t) \times \partial \mathcal{A}
\end{equation}
and we use the terminal condition $u(T,q,x) = x$, that is we set $\gamma =0$. 
Condition \eqref{eq:boundary-cond-u} ensures that $u \approx 0 $ on $\partial A$ for $t < T-\epsilon$ and that at time $T$ boundary and terminal condition are consistent.

If a smooth solution $u$ of the HJB equation \eqref{HJB_bankingsystems} with the boundary condition \eqref{eq:boundary-cond-u} exists, the  optimal trading rate is found by maximizing \eqref{HJB_bankingsystems} with respect to $\nu$, which yields
$$\nu^*(t,q,x)= \frac{\partial_q u }{2 \kappa \partial_x u}(t,x,q).$$
Next we discuss the evolution of the distribution of $\X$. We denote  by $m_t(\ud q,\ud x)$ the distribution of $\X_t$, assuming that banks use the strategy $\nu^*$ and that the contagion term $\overline{\mu}_t$ is equal to  a given deterministic function $\mu(t)$.   For $f\colon \A \to \R$  we use the notation
$\langle m_t,f\rangle =  \int_\mathcal{A} f(q,x) m_t(\ud q,\ud x )$. For a function  $f$ in the domain of $\mathcal{L}^{\mathbf{X}}$ with  $f=0$  on  $\partial \A$, the weak  form of the forward equation for $m_t(\ud q,\ud x)$  is 
$$ \partial_t \langle m_t , f \rangle = \langle m_t, \mathcal{L}^{\mathbf{X}} f\rangle\,  . $$
Partial integration and the  boundary condition  $m(t,q,x) \equiv 0 $ on $\partial \A$ give the forward equation   for the density $m(t,q,x)$
$
\partial_t m (t,q,x)= \big(\mathcal{L}^{\mathbf{X}}\big)^* m(t,q,x)\,,
$
where $\big(\mathcal{L}^{\mathbf{X}}\big)^*$ is the \emph{adjoint} operator  of  $\mathcal{L}^{\mathbf{X}}$.
More explicitly,   
\begin{equation}\label{eq:forward-eq-full}
0= \partial_t m - \frac{1}{2} \sigma_Q^2 \partial_q^2 m - \frac{1}{2} \left( \sigma_A^2 + \sigma_S^2q^2 \right) \partial_x^2 m + \partial_q \left( \nu^* m \right) + \partial_x \left( \big(q( \mu_{\text{ex}} + \alpha \mu(t) )-\kappa ({\nu^*})^2\big ) m \right).
\end{equation}
with initial condition $m(0,\cdot) = m_0$ for a given initial density $m_0$ on $\mathcal{A}$.  

In the case with capital constraints, the assumption that $\overline{\mu}_t$ is the rate of change in the average amount of risky assets in the banking system leads to the following formal definition
\begin{equation}\label{eq:interaction-term-regulated}
 \overline{\mu}_t  = \partial_t \langle m_t,q \rangle = \partial_t \int_\mathcal{A} q \,m_t(\ud x,\ud q)  \,,
 \end{equation}
assuming of course  that this  derivative exists.\footnote{This is not straightforward, see the discussion in Section \ref{subsec:mathematical}.} 

Summarizing, we get the following system of coupled PDEs for an equilibrium of  the  MFGC.
\begin{small}\begin{equation}\label{regulated_PDEsystem}
\left\{
\begin{aligned} 
   0 &= \partial_t u + q(\mu_{\text{ex}} + \alpha \mu(t) ) \partial_x u + \frac{1}{2} \sigma_Q^2 \partial_q^2 u
     + \frac{1}{2} \left(\sigma_A^2 + \sigma_S^2 q^2 \right) \partial_x^2 u
     +  \frac{\left(\partial_q u\right)^2 }{4 \kappa \partial_x u}   &\text{(HJB)}  \\
  0 &=\partial_t m - \frac{1}{2} \sigma_Q^2 \partial_q^2 m - \frac{1}{2} \left( \sigma_A^2 + \sigma_S^2q^2 \right) \partial_x^2 m + \partial_q \left( \frac{\partial_q u }{2 \kappa \partial_x u} m \right) \\
    &+ \partial_x \left( \left(q(\mu_{\text{ex}} + \alpha \mu(t) )- \frac{(\partial_q u)^2 }{4 \kappa  (\partial_x u)^2}\right) m \right)   &\text{(forward)} \\
  \mu(t)  &= \partial_t \langle m_t,q \rangle  &\text{(equilibrium)}\\
  u(T,q,x) &= x   &\text{(terminal)} \\
  m(0,q,x)  &= m_0(x,q)  &\text{(initial)} \\
  u(t,q,x)   &=k(t)\cdot (\beta q+c) \text{ on } \mathcal{A}^c  &\text{(boundary)} \\
  m(t,q,x)   & = 0 \text{ on } \mathcal{A}^c   &\text{(boundary).}
\end{aligned}
\right.
\end{equation}
\end{small}

\subsection{Discussion} Next we describe qualitative properties of solutions under the assumption that a classical solution to \eqref{regulated_PDEsystem} exists. Moreover, we discuss some of the  challenges that arise if one wants to establish mathematical results  on existence and uniqueness of solutions to the system (this issue is not addressed in the present paper).  In Section~\ref{sec:numerical_analysis}  we use numerical methods to study  solutions to the MFGC equations and their dependence on model parameters. 

\subsubsection{Contagion term.} \label{subsec:contagion_term} We  begin  with a discussion of the contagion term $\overline{\mu}_t =\partial_t \langle m_t , q \rangle$.  Intuitively there are two source for  contagion: first, the average selling rate of the active banks  at $t$ \emph{and} second the amount of assets that are liquidated as banks reach the liquidation boundary $\partial \mathcal{A}$. We now  give a mathematical derivation of this  decomposition. 
Suppose that $m(t,q,x)$ is a classical solution of the forward  equation that decays exponentially as $\vert{\mathbf{x}} \vert\to \infty$, and denote by $\big(\mathcal{L}^{\mathbf{X}}\big)^*$ the \emph{adjoint} operator  of  $\mathcal{L}^{\mathbf{X}}$. We  get from the forward equation and  Green's formula that
 $$
 \partial_t \langle m_t , q \rangle = \int_\mathcal{A} q \, \big(\mathcal{L}^{\mathbf{X}}\big)^* m(t,q,x) \,\ud q \ud x = \int_\mathcal{A} m(t,q,x )
  \mathcal{L}^{\mathbf{X}} q \, \ud q \ud x +  \int_{\partial \mathcal{A}} q \partial_{\mathbf{M}} m(t,x,q) \Gamma(\ud x,\ud q )\,,
$$
where $\Gamma(\ud x,\ud q )$ is  the surface element of $\partial \mathcal{A}$ and $\partial_{\mathbf{M}}$ denotes  differentiation  in  direction of the so-called \emph{co-normal $\mathbf{M}$.}\footnote{In principle Greens formula holds only for compact domains, but we can extend this to our case using the assumed exponential decay of $m(t,\mathbf{x})$.}
Using the definition of the co-normal we can give an explicit expression for $\partial_{\mathbf{M}} m(t,q,x)$. Let $\mathbf{n}=( n_1,n_2)'$ be the   outer normal to $\partial \mathcal{A}$. Then
$ \partial_{\mathbf{M}} = \frac{1}{2} \left (\sigma_Q^2 n_1 \partial_Q + (\sigma_A^2 + q^2 \sigma_S^2) n_2  \partial_X\right ).$

Recall that  $ \mathcal{L}^{\mathbf{X}} q = \nu^*(t,\mathbf{x})$. Hence we get, as $\overline {\mu}_t = \partial_t \langle m_t , q \rangle$ by definition, 
\begin{equation} \label{eq:decomp-interaction-term}
\overline{\mu}_t= \langle m_t,\nu^* \rangle + \frac{1}{2} \int_{\partial \mathcal{A}} q \left (\sigma_Q^2 n_1 \partial_Q m(t,q,x) + (\sigma_A^2 + q^2 \sigma_S^2) n_2  \partial_X m(t,q,x)\right ) \ud  \Gamma(\ud x , \ud q )\,. 
\end{equation}
The first  term is the  average trading rate of the active banks, similarly as in  the unregulated case.  The second term gives a mathematical formula for the instantaneous amount of assets that are liquidated as  banks  reach the liquidation boundary $\partial \A$. The form of this term is quite intuitive: it depends 
depends on the  volatility of $Q$ and ${X}$ and on the gradient $\nabla m(t,\mathbf{x})$ on $\partial\A$.  Since $m(t,\mathbf{x}) =0$  for $\mathbf{x} \in \partial \A$,   the partial derivatives of $m$ at the boundary point $\mathbf{x}$ are  a  measure for the mass of the banks  close to the liquidation boundary at $\mathbf{x}$. Hence the second term is large if there are comparatively many banks close to the liquidation boundary and  if the volatility of the state process is large.

\subsubsection{Mathematical challenges.}\label{subsec:mathematical} Next we discuss mathematical challenges arising in the analysis of the system \eqref{regulated_PDEsystem}. We begin with the existence of a smooth density of $\mathbf{X}$. If we combine the forward equation for $m$ and the equilibrium condition we get the following nonlinear differential equation for $m$
\begin{align}
    \partial_t m   &- \frac{1}{2} \sigma_Q^2 \partial_q^2 m - \frac{1}{2} \left( \sigma_A^2 + \sigma_S^2q^2 \right) \partial_x^2 m + \partial_q \left( \frac{\partial_q u }{2 \kappa \partial_x u} m \right) \\
    &+ \partial_x \left( \left(q(\mu_{\text{ex}} + \alpha \partial_t \langle m_t , q \rangle  )- \frac{(\partial_q u)^2 }{4 \kappa  (\partial_x u)^2}\right) m \right) = 0 
\end{align}
This is the so-called McKean-Vlasov equation for $m$. The existence of a smooth solution to this equation is not guaranteed. The main problem is the contagion term:  if the weight $\alpha$  of the contagion term is relatively large or if many banks are close to the liquidation boundary, the feedback effects due to contagion can lead to a \emph{liquidation cascade} where a substantial part of the banking system is  liquidated simultaneously, so that the mapping $t \mapsto \langle m_t,q\rangle$ has a jump.  In the terminology of \citet{bib:nadtochiy-shkolnikov-19} this constitutes  a \emph{systemic risk event}. In a slightly   simpler setting where $X$ is an uncontrolled  one-dimensional process, the existence of solutions to the McKean Vlasov equation  is studied in detail by \citet{bib:delarue-et-al-15} and   \citet{bib:hambly-ledger-sojmark-19}, see also \citet{bib:cuchiero-rigger-svaluto-22}. Based on the  results obtained in these papers  we conjecture that for $\alpha$ sufficiently large there will be a systemic risk event, whereas for $\alpha$ sufficiently small the McKean-Vlasov equation has a solution; this is also in line with the findings from our numerical experiments.  Note however, that our setup is more complicated than the models from \cite{bib:delarue-et-al-15} or \cite{bib:hambly-ledger-sojmark-19}  and that  the parameters $\mu_{\text{ex}}$ and $\kappa$ and the volatility of $X$ and $Q$  play a role as well.

Finally, we comment on existence and uniqueness of an equilibrium for the MFGC, assuming that the McKean Vlasov equation has a solution.  Here we expect positive results in two cases: a) if  the initial distribution $m_0$ has  very little mass close to the liquidation boundary (i.e. if the banking system is well capitalized) the system behaves essentially like the unregulated system for which we have existence results; b) if $\alpha, \tfrac{1}{\kappa}$ and $T$ are not too large we expect a solution to exist due general results on the small-time asymptotics for MFGs.   These conjectures are supported by our numerical experiments,  a formal analysis is however relegated to future research.

\subsection{Numerical Methods for the MFGC}
In order to solve the coupled PDE system resulting from our MFGC numerically, it is necessary to discretize the respective equations. This can be done via finite difference schemes. 
To solve the resulting discrete system, we use an iterative scheme that consists of Picard iterations. Loosely speaking, one starts with a guess $m^{(0)} $ for the flow of measures and one computes the associated contagion term $\overline{\mu}^{(0)}$. Then one  computes the corresponding solution $u^{(1)}$ and the corresponding strategy $\Nu^{(1)}$ of the HJB equation backward in time, using $\overline{\mu}^{(0)}$ as  input,  and one  determines  the dynamics of the corresponding state process $\mathbf{X}^{(1)} := \mathbf{X}^{\Nu^{(1)}}$.   The measure flow  $m^{(1)}$ is then given as  solution of the forward equation  for $\mathbf{X}^{(1)}$. From this one  computes $\overline{\mu}^{(1)}$, then  $u^{(2)}$, and so on,   until some convergence criterion is met.
Refinements  of this approach are discussed in \citet{bib:achdou-20}, and in \citet{bib:achdou-kobeissi-21}. 

In Appendix~\ref{app:numerics}  we present details of our numerical  methodology: we explain how to discretize the system \eqref{regulated_PDEsystem} and we give pseudo-code that explains how we implement the  Picard iteration. To test our implementation  we  compared  the theoretical solution for the unregulated case  derived in Section \ref{sec:unregulated} to  the numerical solution obtained via our implementation of the Picard iteration.  The errors obtained were  very small, so that we feel confident to apply the method also to the case with capital constraints.

\section{Numerical experiments for the case with capital constraints}\label{sec:numerical_analysis}

In this section we report results from numerical experiments. In these experiments  we study how  capital constraints affect the trading rate of individual banks and  the stability of the banking system. Moreover, we study the effectiveness of two macroprudential risk management policies, namely  (i) increasing the  capitalisation of the banking system  and (ii)  improving the  resolution mechanism  for  banks violating  the capital constraints.

For the numerical solution we used $N_k=1000$ time steps, $N_Q=50$ steps in $q$-direction and  $N_X=150$ steps in $x$ direction. We fix the parameters $\gamma=0$, $\sigma_Q=1.4$, $\sigma_S=2$, $\sigma_A=0.1$, $\beta=3$, $c=5$ and $T=1$. The remaining parameter values are reported in Table~\ref{tab:parameters}; these values  vary across  experiments in order to best illustrate certain economic effects. 


\begin{table}[ht]
    \centering
    \begin{tabular}{l cc c c c c c}
        Scenario & $m_0$ && $\alpha$ & $\kappa$ & $\mu_\text{ex}$ & $\alpha_\text{active}$ & $\alpha_\text{liq}$ \\ \hline
        1 & $\mathcal{N} \left( \begin{pmatrix}5\\60\end{pmatrix}, \begin{pmatrix}0.1&0\\0&15\end{pmatrix}\right)$ && $1$ & $20$ & $+1.6$ & & \\ \hline
        2 & $ \mathcal{N} \left( \begin{pmatrix}5\\60\end{pmatrix}, \begin{pmatrix}0.1&0\\0&15\end{pmatrix}\right)$ && $1$ & $20$ & $-1.6$ & & \\ \hline
        3 & $ \mathcal{N} \left( \begin{pmatrix}5\\70\end{pmatrix}, \begin{pmatrix}0.1&0\\0&15\end{pmatrix}\right)$ && $1$ & $20$ & $-1.6$ & & \\ \hline
        4 & $ \mathcal{N} \left( \begin{pmatrix}5\\60\end{pmatrix}, \begin{pmatrix}0.1&0\\0&15\end{pmatrix}\right)$ &&  & $20$ & $-1.6$ & $0.8$ & $0.2$ \\ \hline
    \end{tabular}
    \vspace{0.5cm}
    \caption{Parameter values for numerical experiments }
    \label{tab:parameters}
\end{table}

\subsection{Properties of the optimal trading rate and the value function}
\begin{figure}[h!]
    \centering
    \includegraphics[scale=0.18]{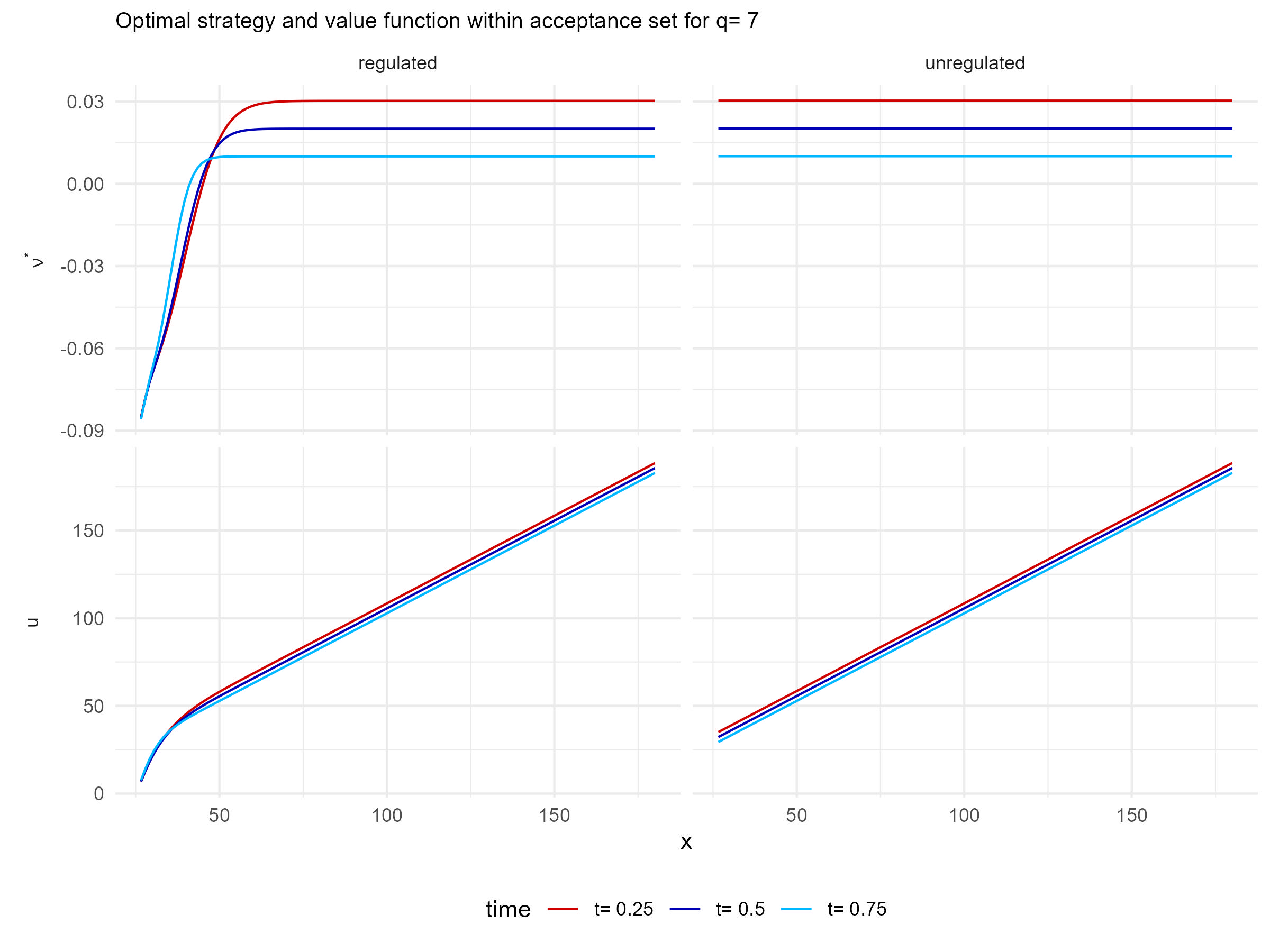}
    \caption{Graph of  the optimal strategy (top) and of the value function (bottom) for fixed $q$ and several time points for  $\mu_\text{ex}=1.6$. Left the case with boundary conditions; right the unregulated case. Parameters as in Scenario 1 of Table~\ref{tab:parameters}. }
    \label{fig:u_nu1}
\end{figure}

In Figure~\ref{fig:u_nu1} we plot sections of the  optimal strategy $\nu^*(t,q,\cdot)$ and of the value function value function $u(t,q,\cdot)$ for varying $x$ and fixed $q=7$ for  various $t$. The other parameters are given in the first line of Table~\ref{tab:parameters}; in particular we assume that   $\mu_\text{ex} =1.6$, so that in the unconstrained case $\nu^* >0$.  The left panel corresponds to the case with capital constraints, the right panel gives the solution in the unregulated case. We see that in the presence of capital constraints, for $x$ close to the liquidation boundary (for $q=7$, $c=5$ and $\beta =3$ at $x=26$) the value function is concave in $x$. The optimal trading rate displays an interesting behavior: for $x$ large it is equal to the optimal trading rate in the unconstrained case and thus constant; as $x$ decreases $\nu^*$ decreases substantially as the bank wants to reduce its inventory to avoid liquidation.

\begin{figure}[h!]
    \centering
    \includegraphics[scale=0.18]{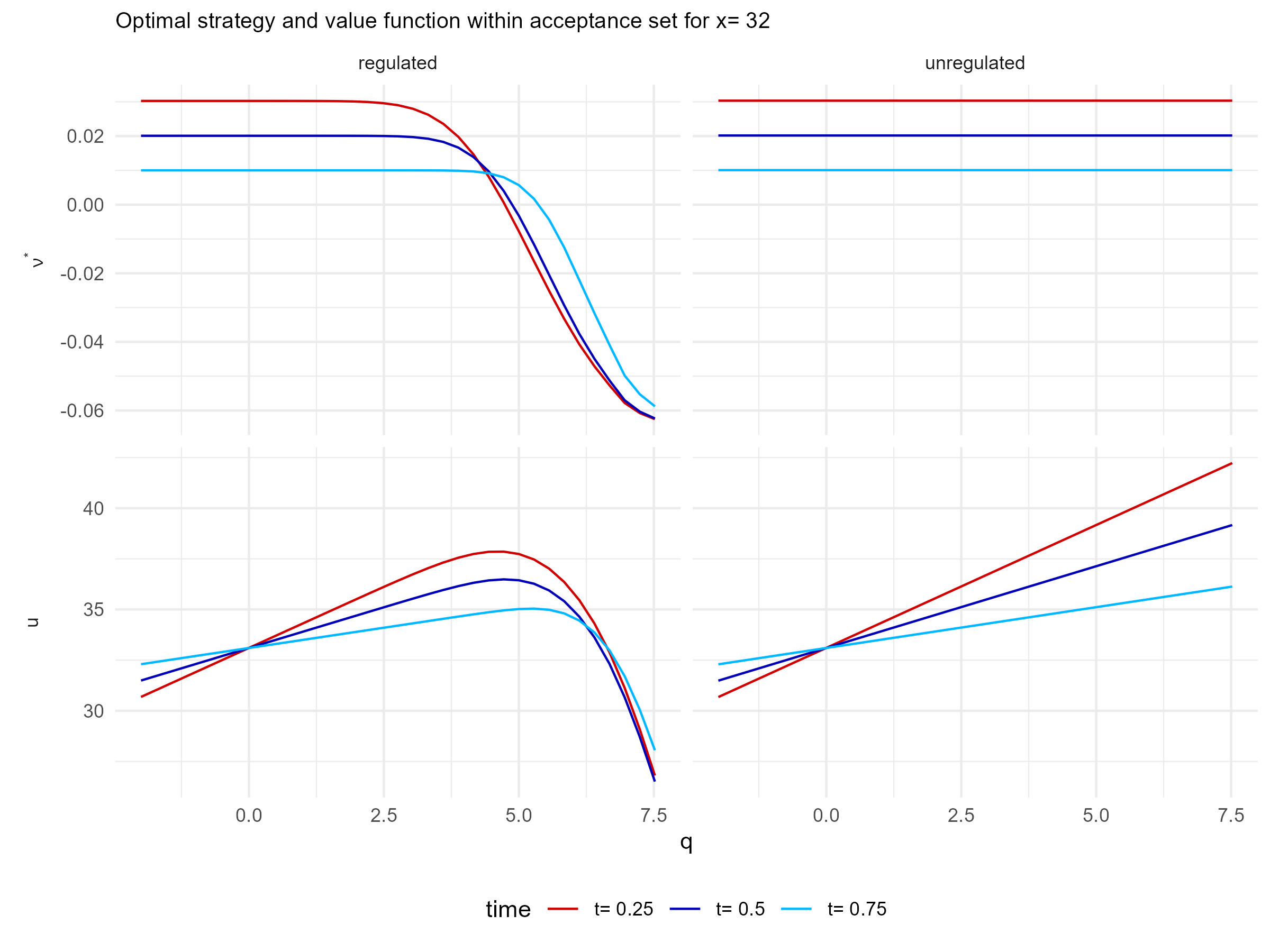}
    \caption{Graph of  the optimal strategy (top) and of the value function (bottom)  for fixed $x$ and several time points for $\mu_\text{ex}=1.6$. Left the case with boundary conditions; right the unregulated case. Parameters as in Scenario 1 of Table~\ref{tab:parameters}. }
    \label{fig:u_nu2}
\end{figure}

In Figure~\ref{fig:u_nu2} we plot sections of the  optimal strategy $\nu^*(t,\cdot,x)$ and of the value function value function $u(t,\cdot,x)$ for varying $q$ and fixed $x=32$ for  various $t$, again for   $\mu_\text{ex} =1.6$. If $q$ is far away from the liquidation boundary, strategy and value function coincide in the regulated and in the unregulated case. We see that $q$ close to the liquidation boundary banks are  deleveraging to avoid liquidation.  Moreover, for $q$ close to the boundary the value function is  decreasing in $q$ whereas in the  case without capital constraints it is increasing throughout (since in that case  a higher inventory level means higher expected profits as $\mu_\text{ex} >0$.  

\subsection{Stability of the banking system}

Next we study the impact of capital constraints on the stability of the banking system. The parameters used are given in scenario 2 of Table~\ref{tab:parameters}.  This  parameter setting corresponding to a \emph{recession scenario} where the assets of all banks trend downward ($\mu_\text{ex}=-1.6$) and where at $t=0$ a large fraction  of the banking system is  close to the liquidation boundary,  due to the low value for the  mean  of the initial  distribution of banks' equity ($\mathbb{E}(X_0)=60$).   Note that the value for $\kappa$ is quite large so that that the  trading rate is low; this corresponds to the case where the tradable asset is relatively illiquid.

\begin{figure}
    \centering
    \includegraphics[scale=0.5]{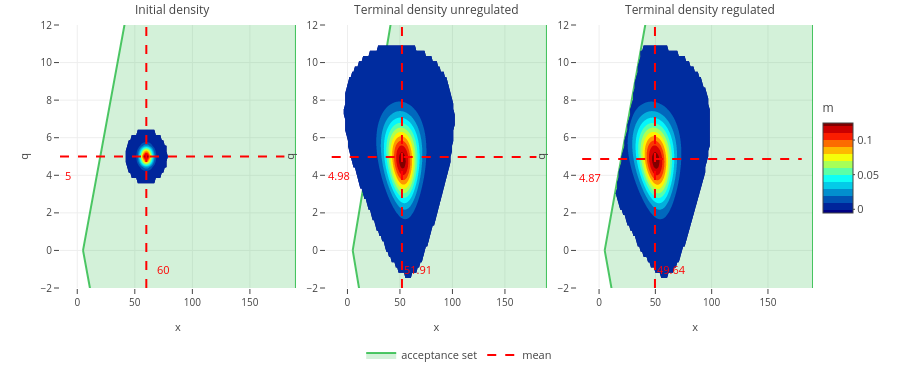}
    \caption{Contour plot of the population density (mean field) at $t=0$  and at $t=T$ for the unregulated case (left) and the regulated case (right); acceptance region $\mathcal{A}$ in green,  parameters as in Scenario~2 of Table~\ref{tab:parameters}.}
    \label{fig:density_contours}
\end{figure}

Figure~\ref{fig:density_contours} shows contour plots of the density $m(t,q,x)$ at $t=0$ and $t=T$.   Due to the negative drift $\mu_\text{ex}=-1.6$ the density curve is transported  to  the left over time (the average  equity value  decreases), as we would  expect in a recession scenario. We observe that in the unregulated case  this move is more pronounced.

\begin{figure}[h!]
    \centering
    \includegraphics[scale=0.18]{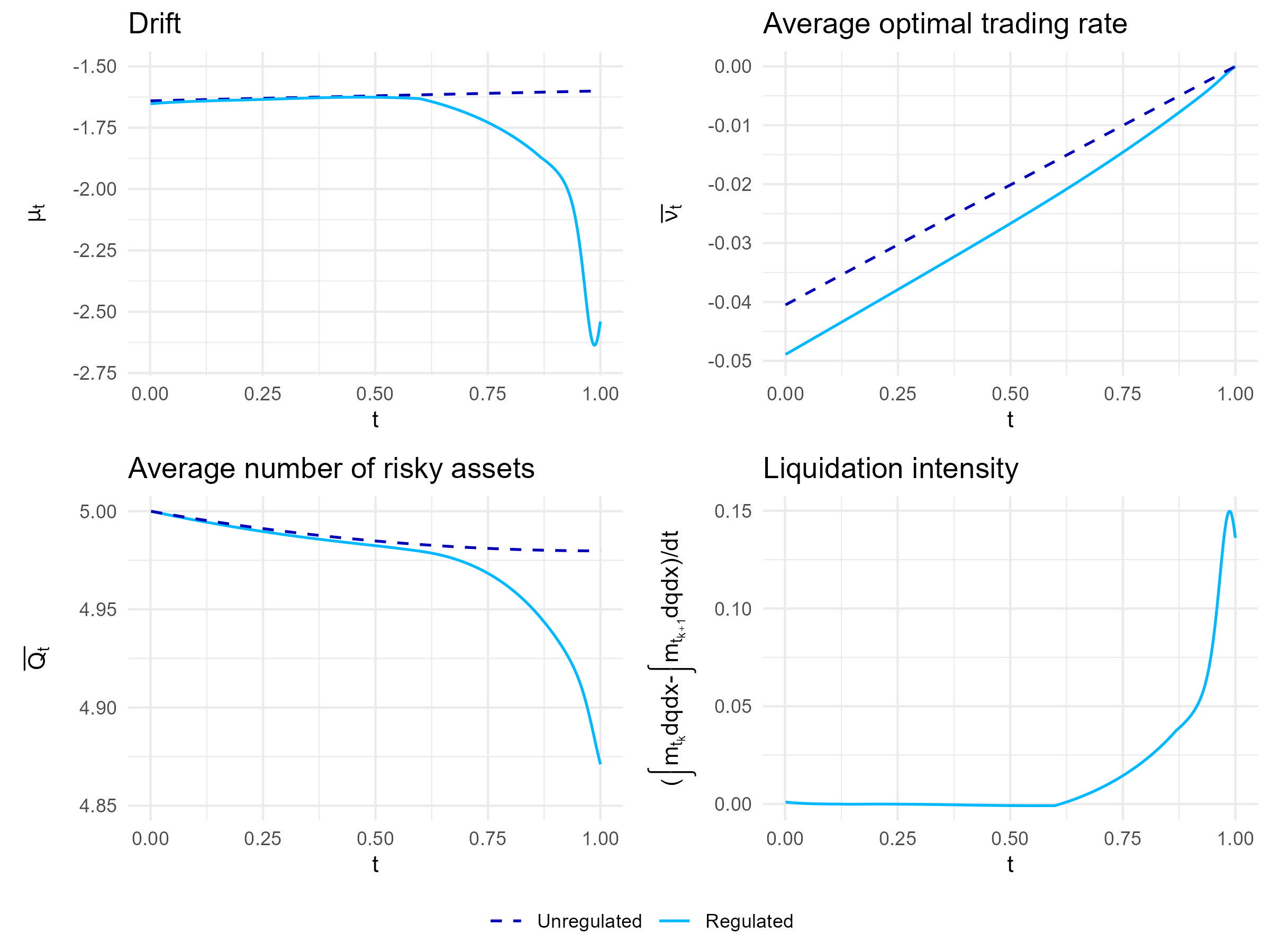}
    \caption{Summary of the banking system in a recession scenario. Parameters as in Scenario 2 of Table~\ref{tab:parameters}. Note in particular the spike in the liquidation intensity at $t=0.9$. } 
    \label{fig:summary-1}
\end{figure}

Figure~\ref{fig:summary-1} presents a summary of the evolution of the banking system. We observe that  there is a strong spike in the liquidation intensity (the change in the proportion of liquidated banks per unit of time)  and a strong decrease in $\overline{\mu}_t$ at $t \approx 0.9$, that is  the system is close to a  systemic crisis.  In fact, if we increase  the parameter   $\alpha$,   the Picard iterations cease  to  converge. Further,  the plots show that in the regulated case the risk bearing capacity of the banking system (the average number of risky assets held by the system) and the mean book value of equity is lower than in the case with capital constraints.


\subsection{Macroprudential policy measures}

Finally, we analyze the impact of two macroprudential policy measures. 

\subsubsection{Increasing capitalization.} 
\begin{figure}[h!]
    \centering
    \includegraphics[scale=0.18]{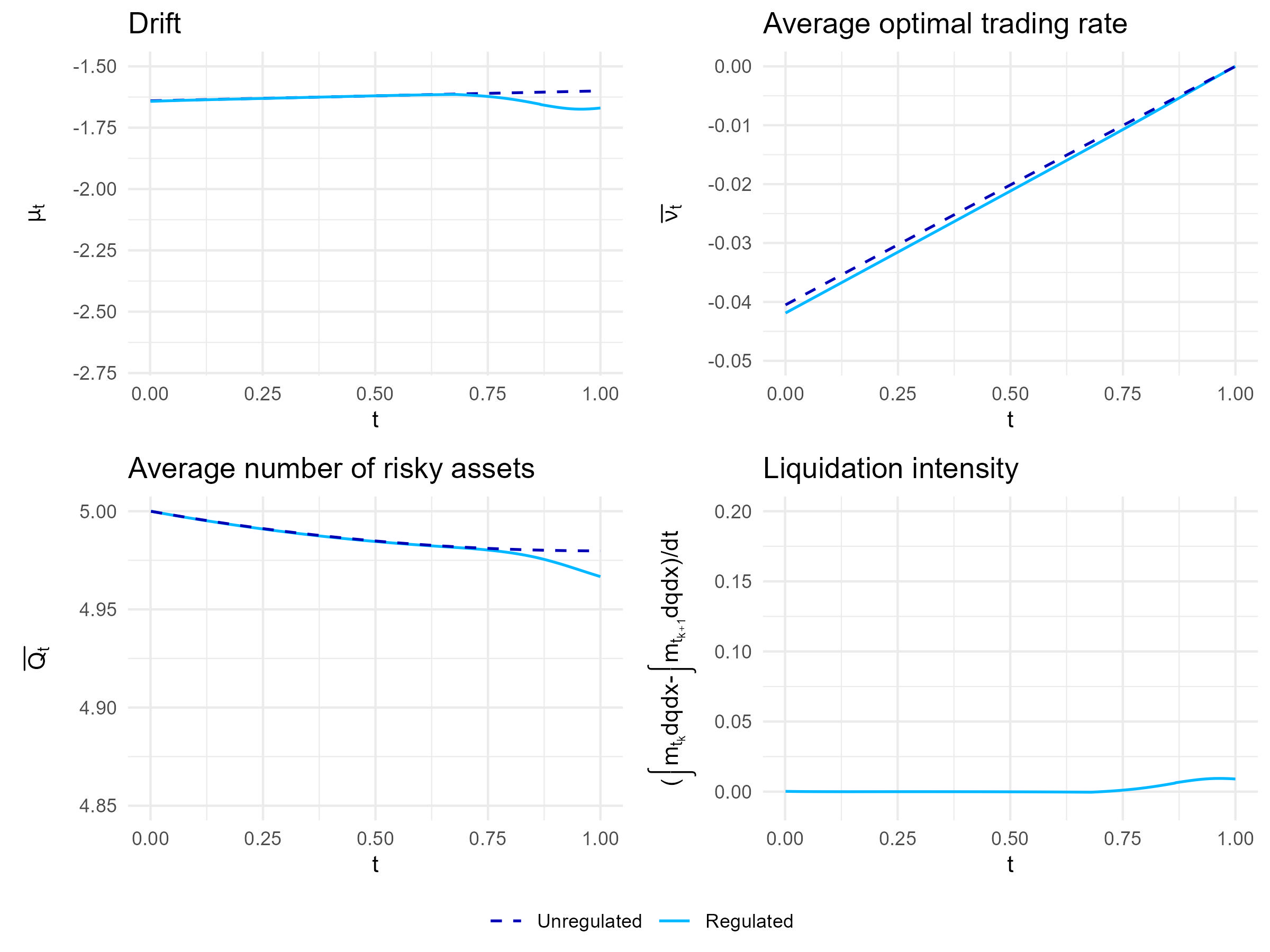}
    \caption{Summary of the banking system with higher initial capital ($\bE[X_0] = 70$), other parameters as in Figure~\ref{fig:summary-1}. Parameters as in Scenario 3 of Table~\ref{tab:parameters}.}
    \label{fig:summary-2}
\end{figure}

It is often argued that sufficient amount of equity capital in the banking system helps to stabilize the system, see for instance \citet{bib:admati-hellwig-13} or \citet{bib:hanson-kashyap-stein-11}. 
We therefore study how a higher level of initial equity affects the initial distribution of the system. In Figure~\ref{fig:summary-2} we plot the banking system for the same parameters as in Figure~\ref{fig:summary-1} except that we now assume that the mean of the initial equity distribution is $\bE(X_0)=70$ (scenario 3 of Table~\ref{tab:parameters}) and thus substantially higher than in Figure~\ref{fig:summary-1}. We observe that the behaviour of the system is very similar to the unregulated case, in particular the spike in the liquidation intensity at $t=0.9$ has almost disappeared. This clearly supports regulatory efforts to ensure that banking systems are well capitalized.


\subsubsection*{Improving the resolution mechanism.} 
\begin{figure}[h!]
    \centering
\includegraphics[scale=.18]{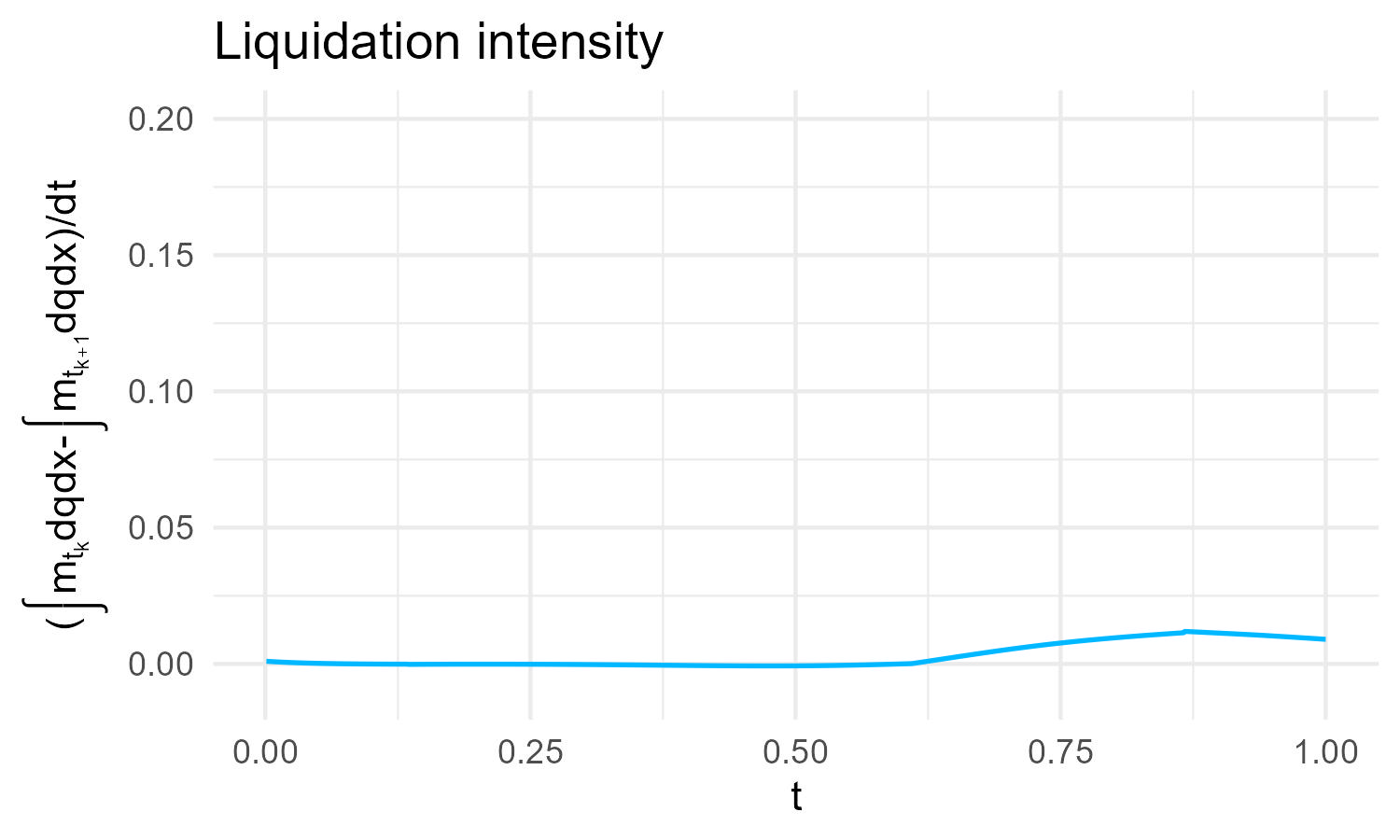}
    \caption{Liquidation intensity with smaller liquidation impact parameter $\alpha_\text{liq}$. Parameters are given in Scenario 4 of Table~\ref{tab:parameters}.}
    \label{fig:liquidation}
\end{figure}

The relatively low values for the average trading rate suggest that a systemic risk event is mostly  due to the automatic and immediate  liquidation of banks upon  violation of  the capital constraints  and less due to the trading behavior of the active banks. This suggests that  financial stability is enhanced if banks violating  the  capital constraints are resolved in a way that mitigates  price-mediated contagion, for instance by parking the assets of these banks in a special purpose vehicle that is unwound only gradually over time.  
A simple way to test this conjecture in our framework is to work with a coefficient $\alpha_\text{active}$ for the impact of the trading of active banks and with a smaller coefficient   $\alpha_{\text{liq}}$ for the impact of the  liquidation. In Figure~\ref{fig:liquidation} we let $\alpha_\text{active} =0.8$ and $\alpha_{\text{liq}} =0.2$ (scenario 4 of Table~\ref{tab:parameters}). Figure ~\ref{fig:liquidation} shows the corresponding liquidation intensity. 

\section{Conclusion}
In this paper we have studied the impact of regulatory capital constraints on fire sales and financial stability in a large banking system using a mean field game
of control (MFGC) model. In our setting  banks  adjust their holdings of a risky asset via trading strategies with finite trading rate.  Moreover, a bank is liquidated if it violates a stylized regulatory capital constraint. We assumed that the drift of the asset value is affected by the average change in the position of the  banks in the system, which creates price mediated contagion.  Numerical methods were used to study properties of the banking system. We found that for high enough values of the contagion parameter a banking system that is not very well capitalized  may experience  a liquidation cascade or systemic risk event, where many banks default more or less simultaneously.
These findings  lends formal support to the conjecture from the regulatory literature  that risk capital constraints may cause price-mediated contagion. They also support   calls for macroprudential approaches to bank regulation (essentially regulatory efforts  trying to mitigate the external effects of rapid deleveraging, see \citet{bib:hanson-kashyap-stein-11}) in addition to standard capital constraints.  Most importantly, regulators should ensure that the banking system is sufficiently well capitalized, as in that case the occurrence of a liquidation cascade is very unlikely. Moreover, it might help to consider improved resolution mechanisms for banks that violate the capital constraints and to park their  assets  in a special purpose vehicle that is unwound  gradually over time.

%


\subsection*{Acknowledgements}
We would like to thank Yves Achdou for his helpful comments regarding the numerical solution of our PDE system.

\clearpage
\appendix

\section{Details on the numerical implementation}
\label{app:numerics}
\subsection{Overview and Notation}
Our goal is to numerically solve the coupled PDE system \eqref{regulated_PDEsystem} by first discretizing it and then using Picard iterations, where the value function and density describing the system are updated until convergence. The main challenges in solving the system numerically are (i) the coupling between the HJB and the forward equation, (ii) the fact that the HJB equation goes backward in time while the PDE describing the evolution of the density goes forward in time, (iii) the natural requirement for the density to be nonnegative and mass-preserving, and (iv) the nonlinearity of our system with respect to various partial derivatives.  One common approach from literature to overcome the first three challenges is to use a special finite difference scheme consisting of a combination of right- and left-sided differences combined with Picard iterations until convergence of the system. We will shortly describe how we tackle the last challenge mentioned that arises from the structure of our PDE system. For completeness, let us start by introducing some standard notation needed for the discretization of the system. 

For  positive integers $N_T,N_Q$,$N_X$, we define the time step size as $\Delta t = \frac{T}{N_T}$ and the step sized related to the state variables $Q$ and $X$ as $\Delta q = \frac{Q_{\text{max}}-Q_{\text{min}}}{N_Q}$ and $\Delta x = \frac{X_{\text{max}}-X_{\text{min}}}{N_X}$, respectively. The set of discrete time steps on our grid is then $\mathfrak{T}=\{ t_k=k\Delta t, k=0,\dots,N_T \}$ and the grid corresponding to the state variable is $\mathfrak{H}= \{h_{ij}= (q_i,x_j)=(Q_\text{min} + i \Delta q, X_\text{min} + j \Delta x) , i=0,\dots,N_Q, j= 0,\dots,N_X\}$. We aim to approximate $u(t_k,h_{i,j})$ and $m(t_k,h_{i,j})$ by $u_{i,j}^k$ and $m_{i,j}^k$, through solving the discrete approximations of the coupled PDE system. We define the following finite difference operators for some function $y:\mathfrak{T} \times \mathfrak{H} \to \mathbb{R}$ 
\begin{align*}
    &D_t y_{i,j}^k = \frac{y_{i,j}^{k+1}-y_{i,j}^k}{\Delta t} ,  & \text{ [Discrete time derivative]} \\
    &D_q y_{i,j}^k = \frac{y_{i+1,j}^k-y_{i-1,j}^k}{2 \Delta q} , & \text{ [Central difference operator in $q$]} \\
    & D_q^R y_{i,j}^k = \frac{y_{i+1,j}^k-y_{i,j}^k}{\Delta q} , & \text{ [Right difference operator in $q$]} \\
    & D_q^L y_{i,j}^k = \frac{y_{i,j}^k-y_{i-1,j}^k}{\Delta q} , & \text{ [Left difference operator in $q$]} \\
    &D_x y_{i,j}^k = \frac{y_{i,j+1}^k-y_{i,j-1}^k}{2 \Delta x} , & \text{ [Central difference operator in $x$]} \\
    &\Delta_q y_i^k = - \frac{1}{{\Delta q}^2} \left( 2 y_{i,j}^k - y_{i+1,j}^k - y_{i-1,j}^k \right) , & \text{ [Central second order difference in $q$]} \\
    &\Delta_x y_{i,j}^k = - \frac{1}{{\Delta x}^2} \left( 2 y_{i,j}^k - y_{i,j+1}^k - y_{i,j-1}^k \right) , & \text{ [Central second order difference in $x$]} \\
    &\Delta_{qx} y_{i,j}^k = \frac{D_q y_{i,j+1}^k - D_q y_{i,j-1}^k }{2 \Delta x} \\
    &= \frac{y_{i+1,j+1}^k-y_{i-1,j+1}^k - y_{i+1,j-1}^k + y_{i-1,j-1}^k }{ \Delta q \Delta x}. & \text{[Mixed second order difference]}
\end{align*}
By definition of these operators, at boundary nodes the grid needs to be extended by one layer. We extend both the value function and the density linearly, i.e. by assuming $u(q+\Delta q,x)=u(q,x)+(u(q,x)-u(q-\Delta q,x))$, $u(q,x+\Delta x)=u(q,x)+(u(q,x)-u(q,x-\Delta x))$, and $m$ similarly. 
Before descritizing our equation system, we also need to rewrite some terms. Note that $$\partial_q \left( \frac{\partial_q u }{2 \kappa \partial_x u} m \right) = \partial_q \left(\frac{\partial_q u }{2 \kappa \partial_x u} \right)m + \frac{\partial_q u }{2 \kappa \partial_x u} \partial_q m =\frac{1}{2\kappa} \frac{ \partial_q^2 u  \partial_x u - \partial_q u \partial_q \partial_x u}{\left(\partial_x u\right)^2} m + \frac{\partial_q u }{2 \kappa \partial_x u} \partial_q m $$ 
and  
\begin{align*}
    &\partial_x \left( \left(q(\mu_{\text{ex}} + \alpha\mu  )- \frac{(\partial_q u)^2 }{4 \kappa (\partial_x u)^2}\right) m \right) =  q (\mu_{\text{ex}} + \alpha\mu  ) \partial_x m - \partial_x \left( \frac{(\partial_q u)^2 }{4 \kappa  (\partial_x u)^2}\right)m - \frac{(\partial_q u)^2 }{4 \kappa  (\partial_x u)^2} \partial_x m \\
    &= q (\mu_{\text{ex}} + \alpha\mu  ) \partial_x m - \frac{1}{4 \kappa } \frac{\partial_x(\partial_q u)^2 \cdot (\partial_x u)^2 - (\partial_q u)^2 \cdot \partial_x (\partial_x u)^2  }{  (\partial_x u)^4}m - \frac{(\partial_q u)^2 }{4 \kappa  (\partial_x u)^2} \partial_x m \\
    &= q (\mu_{\text{ex}} + \alpha\mu  ) \partial_x m - \frac{1}{4 \kappa } \frac{2\partial_q u \partial_x \partial_q u \cdot (\partial_x u)^2 - (\partial_q u)^2 \cdot  2 \partial_x u \partial_x^2u }{  (\partial_x u)^4}m - \frac{(\partial_q u)^2 }{4 \kappa  (\partial_x u)^2} \partial_x m .
\end{align*}

\subsection{Discretization}
In what follows, with regards to the discretization of the system, we follow the idea proposed by \citet{bib:achdou-20}, who use a mix of right- and left-sided difference operators for the discretization in order to be able to overcome the mentioned challenges (i)-(iii). A numerical Hamiltonian is used that is non-increasing in the right-sided differences and non-decreasing in the left-sided differences of the state variable. 

However, we need to modify this idea due to the complexity of our model. In our PDE system, there are terms whose monotonicity with respect to derivatives in different dimension is not clear, as they depend on derivatives with respect to both $q$ and $x$. We therefore use central difference operators in $x$, but distinguish between left- and right-sided difference operators in the dimension of $q$ in order to overcome challenge (iv). Thereby we can achieve better stability of our algorithm. Specifically, we choose the numerical Hamiltonian such that it is non-increasing in the right-sided difference $D_q^R$ and non-decreasing in the left-sided difference $D_q^L$. \\
We discretize the term $\frac{(\partial_q u)^2}{4\kappa \partial_x u}$ in the HJB as $$\max \left\{ \frac{\left[ \left(D_q^R u_{i,j}^k\right)^- \right]^2}{4 \kappa D_x u_{i,j}^k} , \frac{\left[ \left(D_q^L u_{i,j}^k\right)^+ \right]^2}{4 \kappa D_x u_{i,j}^k} \right\} .$$
For the forward equation, we discretize the term $$\partial_q \left( \frac{\partial_q u }{2 \kappa \partial_x u} m \right) = \frac{1}{2\kappa} \frac{ \partial_q^2 u  \partial_x u - \partial_q u \partial_q \partial_x u}{\left(\partial_x u\right)^2} m + \frac{\partial_q u }{2 \kappa \partial_x u} \partial_q m $$
as \\
$ \frac{1}{2\kappa} \frac{\Delta_q u_{i,j}^k D_x u_{i,j}^k - \left\{ \max \left[ \left( D_q^R u_{i,j}^k \right)^+ \left( \Delta_{qx} u_{i,j}^k \right)^+, \left( D_q^L u_{i,j}^k \right)^- \left( \Delta_{qx} u_{i,j}^k \right)^- \right] + \min\left[ \left( D_q^R u_{i,j}^k \right)^- \left( \Delta_{qx} u_{i,j}^k \right)^+, \left( D_q^L u_{i,j}^k \right)^+ \left( \Delta_{qx} u_{i,j}^k \right)^- \right] \right\} }{\left( 
D_x u_{i,j}^k \right)^2 } m_{i,j}^k $
$+ \frac{1}{2 \kappa D_x u_{i,j}^k } \Bigg[ \max\left\{ (D_q^L u_{i,j}^k)^+(D_q^L m_{i,j}^k)^+,(D_q^R u_{i,j}^k)^-(D_q^R m_{i,j}^k)^-\right\} $ \\
$+ \min\left\{ (D_q^R u_{i,j}^k)^+(D_q^L m_{i,j}^k)^-,(D_q^L u_{i,j}^k)^-(D_q^R m_{i,j}^k)^+ \right\} \Bigg] $ ,\\
and the term
\begin{align*}
    &\partial_x \left( \left(q(\mu_{\text{ex}} + \alpha\mu  )- \frac{(\partial_q u)^2 }{4 \kappa (\partial_x u)^2}\right) m \right) \\
    &= q (\mu_{\text{ex}} + \alpha\mu  ) \partial_x m - \frac{1}{4 \kappa } \frac{2\partial_q u \partial_x \partial_q u \cdot (\partial_x u)^2 - (\partial_q u)^2 \cdot  2 \partial_x u \partial_x^2u }{  (\partial_x u)^4}m - \frac{(\partial_q u)^2 }{4 \kappa  (\partial_x u)^2} \partial_x m,
\end{align*}
as \\
$ q(\mu_{ex} + \alpha \mu^k +\delta) D_x m_{i,j}^k $\\
$- \frac{1}{4 \kappa} \frac{2 \left(D_x u_{i,j}^k\right)^2 \cdot \left\{ \max\left[ \left( D_q^L u_{i,j}^k \right)^+ \left( \Delta_{qx} u_{i,j}^k \right)^+,\left( D_q^R u_{i,j}^k \right)^- \left( \Delta_{qx} u_{i,j}^k \right)^- \right] + \min\left[ \left( D_q^L u_{i,j}^k \right)^- \left( \Delta_{qx} u_{i,j}^k \right)^+,\left( D_q^R u_{i,j}^k \right)^+ \left( \Delta_{qx} u_{i,j}^k \right)^- \right] \right\} }{(D_x u_{i,j}^k)^4} m_{i,j}^k $ \\
$+ \frac{1}{4 \kappa} \frac{\max\left\{ \left[ \left( D_q^R u_{i,j}^k \right)^- \right]^2,\left[ \left( D_q^L u_{i,j}^k \right)^+ \right]^2 \right\} \cdot 2 D_x u_{i,j}^k \Delta_x u_{i,j}^k}{(D_x u_{i,j}^k)^4}  m_{i,j}^k$
$- \max \left\{ \frac{\left[ \left(D_q^L u_{i,j}^k\right)^- \right]^2}{4 \kappa \left( D_x u_{i,j}^k \right)^2} , \frac{\left[ \left(D_q^R u_{i,j}^k\right)^+ \right]^2}{4 \kappa \left(D_x u_{i,j}^k\right)^2 } \right\} D_x m_{i,j}^k$.
To shorten the notation we define 
{\tiny\begin{align*}    
    & F_{ij}(u^{k},m^{k})  :=\frac{1}{2\kappa} \Bigg( \frac{\Delta_q u_{i,j}^k D_x u_{i,j}^k }{\left( 
                     D_x u_{i,j}^k \right)^2 } \\
& \,- \frac{ \max \left[ \left( D_q^R u_{i,j}^k \right)^+ \left( \Delta_{qx} u_{i,j}^k \right)^+, \left( D_q^L u_{i,j}^k \right)^- \left( \Delta_{qx} u_{i,j}^k \right)^- \right] + \min\left[ \left( D_q^R u_{i,j}^k \right)^- \left( \Delta_{qx} u_{i,j}^k \right)^+, \left( D_q^L u_{i,j}^k \right)^+ \left( \Delta_{qx} u_{i,j}^k \right)^- \right] }{\left( 
D_x u_{i,j}^k \right)^2 } \Bigg) m_{i,j}^k \\
                       &\,+ \frac{1}{2 \kappa D_x u_{i,j}^k } \Big[ \max\left\{ (D_q^L u_{i,j}^k)^+(D_q^L m_{i,j}^k)^+,(D_q^R u_{i,j}^k)^-(D_q^R m_{i,j}^k)^-\right\} \\
                      &\, + \min\left\{ (D_q^R u_{i,j}^k)^+(D_q^L m_{i,j}^k)^-,(D_q^L u_{i,j}^k)^-(D_q^R m_{i,j}^k)^+ \right\} \Big]
                       +  q(\mu_{ex} + \alpha \mu^k +\delta) D_x m_{i,j}^k \\
                      &\, -\frac{2 \left(D_x u_{i,j}^k\right)^2 \cdot \left\{ \max\left[ \left( D_q^L u_{i,j}^k \right)^+ \left( \Delta_{qx} u_{i,j}^k \right)^+,\left( D_q^R u_{i,j}^k \right)^- \left( \Delta_{qx} u_{i,j}^k \right)^- \right] + \min\left[ \left( D_q^L u_{i,j}^k \right)^- \left( \Delta_{qx} u_{i,j}^k \right)^+,\left( D_q^R u_{i,j}^k \right)^+ \left( \Delta_{qx} u_{i,j}^k \right)^- \right] \right\} }{ 4\kappa (D_x u_{i,j}^k)^4} m_{i,j}^k  \\
                       &\, +  \frac{\max\left\{ \left[ \left( D_q^R u_{i,j}^k \right)^- \right]^2,\left[ \left( D_q^L u_{i,j}^k \right)^+ \right]^2 \right\} \cdot 2 D_x u_{i,j}^k \Delta_x u_{i,j}^k}{ 4 \kappa (D_x u_{i,j}^k)^4}  m_{i,j}^k
- \max \left\{ \frac{\left[ \left(D_q^L u_{i,j}^k\right)^- \right]^2}{4 \kappa \left( D_x u_{i,j}^k \right)^2} , \frac{\left[ \left(D_q^R u_{i,j}^k\right)^+ \right]^2}{4 \kappa \left(D_x u_{i,j}^k\right)^2 } \right\} D_x m_{i,j}^k ,
\end{align*} }

which leads to the following discretized PDE system.

$$ \begin{cases}
  D_t u_{i,j}^k + q_i(\mu_{\text{ex}} +\alpha\mu^k   ) D_x u_{i,j}^k + \frac{1}{2} \sigma_Q^2 \Delta_q u_{i,j}^k + \frac{1}{2} \left(\sigma_A^2 + \sigma_S^2 q_i^2 \right) \Delta_x u^k_{i,j} \\
  +  \max \left\{ \frac{\left[ \left(D_q^R u_{i,j}^k\right)^- \right]^2}{4 \kappa D_x u_{i,j}^k} , \frac{\left[ \left(D_q^L u_{i,j}^k\right)^+ \right]^2}{4 \kappa D_x u_{i,j}^k} \right\}  =0 , \\
  D_t m_{i,j}^k - \frac{1}{2} \sigma_Q^2 \Delta_q m_{i,j}^k - \frac{1}{2} \left( \sigma_A^2 + \sigma_S^2q^2 \right) \Delta_x m_{i,j}^k +F_{i,j}(u^k,m^k) =0 \\
 \mu^k = D_t \left( \sum_{i } \sum_{j} q_i m_{i,j}^k \Delta x \Delta q \right), \\
  u(T,q_i,x_j)= x_j- \gamma q_i^2 , \\
  m(0,q_i,x_j) = m_0(x_i,q_j) , \\
  u(t_k,q_i,x_j) \equiv k(t_k)\cdot (\beta q_i+c) \text{ on }  \mathcal{A}^c, \\
  m(t_k,q_i,x_j) \equiv 0 \text{ on }  \mathcal{A}^c .
\end{cases} $$

\subsection{Picard Iteration}
The discretized PDE system can now be used to update initial guesses of the value function $u$ (backward in time) and the density $m$ (forward in time) iteratively, until a convergence criterion is met.

We implement these Picard iterations in \texttt{R}. The following pseudo code shows the structure of the algorithm.
           \begingroup
           \setlength{\tabcolsep}{3pt}
           \begin{table}[]
            \begin{tabular}{llllll}
                \multicolumn{5}{l}{Iterative Solvers} \\ \hline
                \multicolumn{5}{l}{\small \textbf{function}  $\mathtt{iterate_u(u,m)}$ } & \scriptsize iteration of u \\
                    &  \multicolumn{4}{l}{\small \textbf{for } $k=N_T-1,\dots,1 $} & \scriptsize backward iteration \\
                         & & \multicolumn{3}{l}{\footnotesize  $l=0$ } &\\
                         & & \multicolumn{3}{l}{\footnotesize  $u^0=u$ } &\\
                         & &  \multicolumn{3}{l}{\footnotesize $\mu^k=D_t \left( \sum_{i } \sum_{j} q_i m_{i,j}^k \Delta x \Delta q \right)$}  & \scriptsize drift \\
                        & & \multicolumn{3}{l}{\small \textbf{while} error $>$ tolerance} &\\
                        & & & \multicolumn{2}{l}{\footnotesize $l = l +1$}  & \scriptsize iteration step \\
                     & & & \multicolumn{2}{l}{\footnotesize $u_{ij}^{k,l}= u_{ij}^{k+1,l} + \Delta t \cdot \Bigg[ q_i(\mu_{\text{ex}} +\alpha\mu^k   ) D_x u_{i,j}^{k,l} + \frac{1}{2} \sigma_Q^2 \Delta_q u_{i,j}^{k,l}  $} & \scriptsize updated value function \\
                     & & & & \multicolumn{1}{l}{\footnotesize $+ \frac{1}{2} \left(\sigma_A^2 + \sigma_S^2 q_i^2 \right) \Delta_x u^{k,l}_{i,j} +  \max \left\{ \frac{\left[ \left(D_q^R u_{i,j}^{k,l}\right)^- \right]^2}{4 \kappa D_x u_{i,j}^{k,l}} , \frac{\left[ \left(D_q^L u_{i,j}^{k,l}\right)^+ \right]^2}{4 \kappa D_x u_{i,j}^{k,l}} \right\}\Bigg]$ }
                      \\
                     & & & \multicolumn{2}{l}{\small \textbf{if} $\text{regulated}=\texttt{TRUE}$ \quad \footnotesize $u_{ij}^{k,l}[\mathcal{A}^c]=k(t_k)\cdot (\beta q_i +c)$ } & \scriptsize boundary condition \\
                     & & & \multicolumn{2}{l}{\footnotesize $\text{error}=\text{mean}( \text{abs}(u_{ij}^{k,l} - u_{ij}^{k,l-1} ) )$ }  & \scriptsize error between guesses  \\
                     \multicolumn{5}{l}{\small \textbf{return} \footnotesize $u$} & \\ 
                     \hline
                     \multicolumn{5}{l}{\small \textbf{function}  $\mathtt{iterate_m(u,m)}$ } & \scriptsize iteration of m \\
                    &  \multicolumn{4}{l}{\small \textbf{for } $k=1,\dots,N_T-1 $} & \scriptsize forward iteration \\
                     & & \multicolumn{3}{l}{\footnotesize $p=0$} &\\
                     & & \multicolumn{3}{l}{\footnotesize $m^0=m$} &\\
                        & & \multicolumn{3}{l}{\small \textbf{while} error $>$ tolerance} &\\    
                     & & & \multicolumn{2}{l}{\footnotesize $\mu^{k,p}=D_t \left( \sum_{i } \sum_{j} q_i m_{i,j}^{k,p} \Delta x \Delta q \right)$}  & \scriptsize drift \\
                     & & & \multicolumn{2}{l}{\footnotesize $p=p+1$}  & \scriptsize iteration step \\
                     & & & \multicolumn{2}{l}{\footnotesize  $m_{ij}^{k+1,p}= m_{ij}^{k,p} - \Delta t 
                      \Big[ - \frac{1}{2} \sigma_Q^2 \Delta_q m_{i,j}^{k,p} - \frac{1}{2} \left( \sigma_A^2 + \sigma_S^2q^2 \right) \Delta_x m_{i,j}^{k,p} + F_{ij}(u^{k,p},m^{k,p}) \Big]
                      $} & \scriptsize updated density
                      \\
                     & & & \multicolumn{2}{l}{\small \textbf{if} $\text{regulated}=\texttt{TRUE}$ \quad \footnotesize $m_{ij}^{k+1,p}[\mathcal{A}^c]=0$ } & \scriptsize boundary condition \\
                     & & & \multicolumn{2}{l}{\footnotesize $\text{error}=\text{mean}( \text{abs}(m_{ij}^{k+1,p} - m_{ij}^{k+1,p-1} ) )$ }  & \scriptsize error between guesses  \\
                     \multicolumn{5}{l}{\small \textbf{return} \footnotesize $m$} & \\ 
                     \hline

                \multicolumn{5}{l}{Picard Iteration} \\ \hline
                \multicolumn{5}{l}{\footnotesize $u=u_T $, $m=m_0$} & \scriptsize initialization \\
                \multicolumn{5}{l}{\footnotesize $n=0$} & \scriptsize iteration step \\
                \multicolumn{5}{l}{\small \textbf{while} error $>$ tolerance}  & \scriptsize iteration on whole grid \\
                & \multicolumn{4}{l}{\footnotesize $n=n+1$} & \scriptsize iteration step  \\
                & \multicolumn{4}{l}{\footnotesize $u=\mathtt{iterate_u(u,m)}$ } & \scriptsize iteration of u  \\
                & \multicolumn{4}{l}{\footnotesize $m= \mathtt{iterate_m(u,m)}$ } & \scriptsize iteration of m \\
                & \multicolumn{4}{l}{\footnotesize $\text{error}= \Big[ \text{mean}( \text{abs}(u_{ij}^{k,n} - u_{ij}^{k,n-1} ) ) + \text{mean}( \text{abs}(m_{ij}^{k,n} - m_{ij}^{k,n-1} ) ) \Big]/2$  } &  \scriptsize error between guesses \\
                     \hline
               \end{tabular}
           \end{table}
           \endgroup \\

\clearpage 

\section{Details on the explicit solution in the unregulated case}
\label{app:explicitSol}

In this section we give explicit expressions for the functions $h_0$ and $h_1$ that arise in the explicit solution for the MFGC in the unconstrained case. We get for $h_1$
\begin{align*}
    h_1 &= 2 \kappa E' + h_2 E \\
    &= 2 \kappa \left[ \frac{ \left(  \mu_{\text{ex}} \alpha + \frac{\alpha\mu_{\text{ex}}\gamma}{\kappa} T -E_0  \frac{\alpha^2 \gamma }{\kappa} \right)  e^{-\frac{\alpha}{2\kappa}t} }{ \left( \alpha^2 -2\gamma \alpha \right) e^{-\frac{\alpha}{2\kappa}T} + 2\gamma \alpha }  - \frac{\mu_{\text{ex}}}{\alpha} \right] \\
    & + \frac{2 \kappa}{T-t+ \frac{\kappa}{\gamma}} \left[ \frac{ E_0 \left( \alpha^2 e^{-\frac{\alpha}{2\kappa}T} + 2 \gamma \alpha \left(  e^{-\frac{\alpha}{2\kappa}t} - e^{-\frac{\alpha}{2\kappa}T}\right) \right) - \left( 2 \kappa \mu_{\text{ex}} + 2 \mu_{\text{ex}} \gamma T\right) \left( e^{-\frac{\alpha}{2\kappa}t} -1 \right)   }{ \left( \alpha^2 -2\gamma \alpha \right) e^{-\frac{\alpha}{2\kappa}T} + 2\gamma \alpha }  - \frac{\mu_{\text{ex}}}{\alpha} t \right] .
\end{align*}
Given $h_1$, we can compute $h_0$ using \eqref{eq:c},
\begin{align*}
    h_0' &= - \frac{h_1^2}{4 \kappa} = - \frac{4 \kappa^2 E'^2 + 4 \kappa h_2 E E' + h_2^2E^2 }{4 \kappa}  = - \kappa E'^2 - \frac{2 \kappa}{T-t+ \frac{\kappa}{\gamma}} E E' - \frac{\kappa}{(T-t+ \frac{\kappa}{\gamma})^2} E^2 \\
    &= - \kappa \left[ \frac{ \left(  \mu_{\text{ex}} \alpha + \frac{\alpha\mu_{\text{ex}}\gamma}{\kappa} T -E_0  \frac{\alpha^2 \gamma }{\kappa} \right)  e^{-\frac{\alpha}{2\kappa}t} }{ \left( \alpha^2 -2\gamma \alpha \right) e^{-\frac{\alpha}{2\kappa}T} + 2\gamma \alpha }  - \frac{\mu_{\text{ex}}}{\alpha} \right]^2 \\
    &- \frac{2 \kappa}{T-t + \frac{\kappa}{\gamma}} \left[ \frac{ E_0 \left( \alpha^2 e^{-\frac{\alpha}{2\kappa}T} + 2 \gamma \alpha \left(  e^{-\frac{\alpha}{2\kappa}t} - e^{-\frac{\alpha}{2\kappa}T}\right) \right) - \left( 2 \kappa \mu_{\text{ex}} + 2 \mu_{\text{ex}} \gamma T\right) \left( e^{-\frac{\alpha}{2\kappa}t} -1 \right)   }{ \left( \alpha^2 -2\gamma \alpha \right) e^{-\frac{\alpha}{2\kappa}T} + 2\gamma \alpha }  - \frac{\mu_{\text{ex}}}{\alpha} t \right] \\
    &\cdot \left[ \frac{ \left(  \mu_{\text{ex}} \alpha + \frac{\alpha\mu_{\text{ex}}\gamma}{\kappa} T -E_0  \frac{\alpha^2 \gamma }{\kappa} \right)  e^{-\frac{\alpha}{2\kappa}t} }{ \left( \alpha^2 -2\gamma \alpha \right) e^{-\frac{\alpha}{2\kappa}T} + 2\gamma \alpha }  - \frac{\mu_{\text{ex}}}{\alpha} \right] \\
    &-  \frac{\kappa}{(T-t+ \frac{\kappa}{\gamma})^2} \left[ \frac{ E_0 \left( \alpha^2 e^{-\frac{\alpha}{2\kappa}T} + 2 \gamma \alpha \left(  e^{-\frac{\alpha}{2\kappa}t} - e^{-\frac{\alpha}{2\kappa}T}\right) \right) - \left( 2 \kappa \mu_{\text{ex}} + 2 \mu_{\text{ex}} \gamma T\right) \left( e^{-\frac{\alpha}{2\kappa}t} -1 \right)   }{ \left( \alpha^2 -2\gamma \alpha \right) e^{-\frac{\alpha}{2\kappa}T} + 2\gamma \alpha }  - \frac{\mu_{\text{ex}}}{\alpha} t \right]^2 \\
    &=: \Omega' + \Gamma' + \Pi'.
\end{align*}
For $h_0$ we obtain 
\[h_0= \Omega + \Gamma + \Pi,\]
where 
\begin{align*}
    \Omega &= - \frac{\kappa}{\alpha} \frac{ \left(- \mu_{\text{ex}}^2 \alpha^2 \kappa - \frac{\alpha^2 \mu_{\text{ex}}^2  \gamma^2}{\kappa}T^2 - E_0^2\frac{\alpha^4 \gamma^2}{\kappa} - 2  \alpha^2 \mu_{\text{ex}}^2 \gamma T + 2 E_0 \alpha^3 \mu_{\text{ex}} \gamma + 2 \frac{\alpha^3 \mu_{\text{ex}} \gamma^2}{\kappa} T E_0 \right) e^{-\frac{\alpha}{\kappa} t} }{ \left( \left( \alpha^2 -2\gamma \alpha \right) e^{-\frac{\alpha}{2\kappa}T} + 2\gamma \alpha \right)^2} \\
    &- 2 \frac{2 \kappa}{\alpha} \frac{  \left(  \mu_{\text{ex}}^2 \kappa + \mu_{\text{ex}}^2\gamma T -E_0  \alpha \gamma \mu_{\text{ex}} \right)  e^{-\frac{\alpha}{2\kappa}t} }{ \left( \alpha^2 -2\gamma \alpha \right) e^{-\frac{\alpha}{2\kappa}T} + 2\gamma \alpha } - \frac{\mu_{\text{ex}}^2 \kappa}{\alpha^2} t + C_1  \\
    &= \frac{ \left( \mu_{\text{ex}}^2 \alpha \kappa^2 + \alpha  \mu_{\text{ex}}^2  \gamma^2T^2 + E_0^2 \alpha^3 \gamma^2 + 2  \alpha \kappa \mu_{\text{ex}}^2 \gamma T - 2 E_0 \alpha^2 \mu_{\text{ex}} \kappa \gamma - 2 \alpha^2 \mu_{\text{ex}} \gamma^2 T E_0 \right) e^{-\frac{\alpha}{\kappa} t} }{ \left( \left( \alpha^2 -2\gamma \alpha \right) e^{-\frac{\alpha}{2\kappa}T} + 2\gamma \alpha \right)^2} \\
    &- 4 \frac{  \left(  \frac{ \mu_{\text{ex}}^2 \kappa^2}{\alpha} + \frac{ \mu_{\text{ex}}^2 \kappa \gamma}{\alpha} T -E_0  \kappa \gamma \mu_{\text{ex}} \right)  e^{-\frac{\alpha}{2\kappa}t} }{ \left( \alpha^2 -2\gamma \alpha \right) e^{-\frac{\alpha}{2\kappa}T} + 2\gamma \alpha } - \frac{\mu_{\text{ex}}^2 \kappa}{\alpha^2} t + C_1,
\end{align*}
\begin{align*}
    \Gamma &= - \Bigg[  \frac{ \left( \left( 4 \gamma \alpha E_0 -  2 \alpha^2 E_0 \right)  e^{-\frac{\alpha}{2 \kappa} T} - \left( 4 \kappa \mu_{\text{ex}} + 4 \mu_{\text{ex}} \gamma T \right) \right) \left(  \mu_{\text{ex}} \alpha \kappa + \alpha\mu_{\text{ex}}\gamma T -E_0  \alpha^2 \gamma \right)   }{ \left( \left( \alpha^2 -2\gamma \alpha \right) e^{-\frac{\alpha}{2\kappa}T} + 2\gamma \alpha \right)^2} \\
    &+ \frac{ 8 E_0 \gamma \mu_{\text{ex}} \kappa - \left( 4 \frac{ \kappa^2 \mu_{\text{ex}}^2}{\alpha} + 4 \frac{ \mu_{\text{ex}}^2 \gamma \kappa }{\alpha} T\right) }{  \left( \alpha^2 -2\gamma \alpha \right) e^{-\frac{\alpha}{2\kappa}T} + 2\gamma \alpha  } \Bigg] 
    e^{- \frac{\alpha (\gamma T + \kappa)}{2 \gamma \kappa}} \operatorname{Ei} \left( \frac{\alpha (\kappa + \gamma (T-t))}{2 \gamma \kappa} \right)  \\ 
    & - \frac{  \left( 4 \kappa \mu_{\text{ex}} +4 \mu_{\text{ex}} \gamma T  - 4 \gamma \alpha E_0 \right)  \left(  \mu_{\text{ex}} \alpha \kappa + \alpha\mu_{\text{ex}}\gamma T -E_0  \alpha^2 \gamma \right)  }{ \left( \left( \alpha^2 -2\gamma \alpha \right) e^{-\frac{\alpha}{2\kappa}T} + 2\gamma \alpha \right)^2} e^{-\frac{\alpha (\gamma T +\kappa )}{\gamma \kappa}} \operatorname{Ei} \left( \frac{\alpha (\kappa + \gamma (T-t)) }{\gamma \kappa} \right) \\
    &- \frac{\left( 2 E_0  \alpha \mu_{\text{ex}} \kappa - 8 E_0 \gamma \mu_{\text{ex}} \kappa \right) e^{-\frac{\alpha}{2\kappa}T} + 4 \frac{ \kappa^2 \mu_{\text{ex}}^2}{\alpha} + 4 \frac{ \mu_{\text{ex}}^2 \gamma \kappa }{\alpha} T  }{  \left( \alpha^2 -2\gamma \alpha \right) e^{-\frac{\alpha}{2\kappa}T} + 2\gamma \alpha  } \log \left( T-t+ \frac{\kappa}{\gamma} \right)
    \\
    &+ \frac{\left(  2 \mu_{\text{ex}}^2 \kappa + 2 \mu_{\text{ex}}^2\gamma T - 2 E_0  \alpha \mu_{\text{ex}} \gamma  \right) }{  \left( \alpha^2 -2\gamma \alpha \right) e^{-\frac{\alpha}{2\kappa}T} + 2\gamma \alpha  } 
    \frac{e^{- \frac{\alpha (\kappa + \gamma (T+t))}{2\gamma \kappa}}}{\alpha \gamma} \left[ 2 \gamma \kappa e^{\frac{\alpha (\kappa + \gamma T) }{2\gamma \kappa}} - \alpha e^{\frac{\alpha t}{2 \kappa}} (\gamma T + \kappa) \operatorname{Ei} \left( \frac{\alpha (\kappa+\gamma(T-t))}{2 \gamma \kappa} \right) \right] \\
    &+  \frac{2 \kappa \mu_{\text{ex}}^2}{\alpha^2} \left(  \frac{\gamma T + \kappa}{\gamma} \log \left( \gamma (T-t) + \kappa \right) + t \right)  + C_2,
\end{align*}
and
\begin{align*}
    \Pi &=  - \frac{ \left( 2 E_0 \gamma \alpha - 2 \kappa \mu_{\text{ex}}  - 2 \mu_{\text{ex}} \gamma T \right)^2 }{ \left( \left( \alpha^2 -2\gamma \alpha \right) e^{-\frac{\alpha}{2\kappa}T} + 2\gamma \alpha \right)^2 } 
    \frac{ e^{-\frac{\alpha(\gamma T + \kappa)}{\gamma \kappa}} \left( \gamma \kappa e^{\frac{\alpha (\gamma (T-t) + \kappa ) }{\gamma \kappa }} - \alpha (\gamma (T-t) + \kappa ) \operatorname{Ei} \left( \frac{\alpha (\gamma (T-t) + \kappa )}{\gamma \kappa} \right)  \right) }{ \gamma (T-t) + \kappa} \\
    &- \frac{  \left( 4 E_0 \gamma \alpha - 4 \kappa \mu_{\text{ex}}  -4 \mu_{\text{ex}} \gamma T \right)  \left(E_0  \alpha^2  - 2 E_0 \gamma \alpha\right) e^{-\frac{\alpha}{2\kappa}T} +  \left( 4 E_0 \gamma \alpha - 4 \kappa \mu_{\text{ex}}  - 4\mu_{\text{ex}} \gamma T \right)  \left( 2 \kappa \mu_{\text{ex}}  + 2 \mu_{\text{ex}} \gamma T \right) }{ \left( \left( \alpha^2 -2\gamma \alpha \right) e^{-\frac{\alpha}{2\kappa}T} + 2\gamma \alpha \right)^2 } 
    \\
    & \cdot \frac{ e^{-\frac{\alpha (\gamma T + \kappa )}{2\gamma \kappa}} \left( 2 \gamma \kappa e^{\frac{\alpha (\gamma (T-t) + \kappa)}{2 \gamma \kappa}} - \alpha (\gamma (T-t) + \kappa ) \operatorname{Ei} \left( \frac{ \alpha (\gamma(T-t)+\kappa)}{2 \gamma \kappa} \right) \right) }{2  (\gamma (T-t)+\kappa )} \\
    &-  \frac{  \left(E_0  \alpha^2  - 2 E_0 \gamma \alpha\right)^2 e^{-\frac{\alpha}{\kappa}T}   + \left(2 \kappa \mu_{\text{ex}}  + 2 \mu_{\text{ex}} \gamma T \right)^2 + 2 \left(E_0  \alpha^2  - 2 E_0 \gamma \alpha\right) e^{-\frac{\alpha}{2\kappa}T}  \left( 2 \kappa \mu_{\text{ex}}  + 2 \mu_{\text{ex}} \gamma T \right)  }{ \left( \left( \alpha^2 -2\gamma \alpha \right) e^{-\frac{\alpha}{2\kappa}T} + 2\gamma \alpha \right)^2 } 
    \frac{\kappa }{T-t+ \frac{\kappa}{\gamma}} \\ 
    & + \frac{  4 E_0 \gamma \mu_{\text{ex}}    - \left( 4 \frac{\kappa \mu_{\text{ex}}^2}{\alpha} + 4 \frac{\mu_{\text{ex}}^2 \gamma}{\alpha} T\right)      }{ \left( \alpha^2 -2\gamma \alpha \right) e^{-\frac{\alpha}{2\kappa}T} + 2\gamma \alpha }  \frac{\kappa}{2\gamma \kappa (\gamma(T-t)+\kappa)}e^{-\frac{\alpha (\kappa + \gamma T)}{2 \gamma \kappa}} \\
    & \cdot \left( - (\gamma (T-t) + \kappa) (\alpha \kappa - 2 \gamma \kappa + \gamma \alpha T)  \operatorname{Ei} \left( \frac{\alpha (\kappa + \gamma (T-t))}{2\gamma \kappa} \right) + 2 \gamma \kappa (\kappa + \gamma T) e^{\frac{\alpha (\kappa + \gamma (T-t))}{2 \gamma \kappa }}\right) \\
    &+ \frac{ 2 E_0  \alpha \mu_{\text{ex}} e^{-\frac{\alpha}{2\kappa}T}  - 4 E_0 \gamma \mu_{\text{ex}}  e^{-\frac{\alpha}{2\kappa}T}  + \left( 4 \frac{\kappa \mu_{\text{ex}}^2}{\alpha} + 4 \frac{\mu_{\text{ex}}^2 \gamma}{\alpha} T\right)    }{ \left( \alpha^2 -2\gamma \alpha \right) e^{-\frac{\alpha}{2\kappa}T} + 2\gamma \alpha }  \frac{\kappa \left[ (\gamma (T-t) + \kappa) \log \left( T-t +\frac{\kappa}{\gamma} \right) + \gamma T + \kappa  \right] }{\gamma (T-t) + \kappa} \\
    &- \frac{\mu_{\text{ex}}^2 \kappa }{\alpha^2} \left[ \frac{(\gamma T + \kappa)^2 }{\gamma^2 \left(T-t + \frac{\kappa}{\gamma}\right)} + \frac{2 (\gamma T + \kappa) \log \left( T-t + \frac{\kappa }{\gamma } \right) }{\gamma } + t \right] + C_3.  
\end{align*}
Note that it is determined only up to a constant $C:= C_1+C_2+C_3$. Recall that $h_0(T)=0$, from which we can determine the constant. $\operatorname{Ei}$ denotes the exponential integral.

\clearpage

\bibliographystyle{plainnat}
\bibliography{Banking.bib}

\end{document}